\documentclass[preprint,showpacs,12pt,floatfix,aps]{revtex4}

\usepackage{graphicx}
\usepackage{latexsym}
\usepackage{amsmath}
\input psfig.sty

\begin{document}

\title{Radiating Gravitational Collapse with an Initial Inhomogeneous Energy 
Density Distribution} 
\author{Pinheiro, G. $^{1}$}
\email{gpinheiro@on.br}
\author{R. Chan $^{2}$}
\email{chan@on.br}
\affiliation{\small $^{1}$ Divis\~ao de Programas de P\'os-Gradua\c c\~ao 
$^{2}$ Coordena\c{c}\~ao de Astronomia e Astrof\'{\i}sica,
Observat\'orio Nacional, Rua General Jos\'e Cristino, 77, S\~ao Crist\'ov\~ao
20921-400, Rio de Janeiro, RJ, Brazil}

\date{18/11/2010}

\begin{abstract}
A new model is proposed to a collapsing star consisting of an initial
inhomogeneous energy density and anisotropic pressure fluid with shear, 
radial heat flow and outgoing
radiation.  In previous papers one of us has always assumed an initial star with
homogeneous energy density.
The aim of this work is to generalize the previous models by introducing an
initial inhomogeneous energy density and compare it to the initial
homogeneous energy density collapse model.  We will show the differences 
between these models in the evolution of all physical quantities
that characterizes the gravitational collapse.
The behavior of the energy density, pressure, mass, luminosity and the
effective adiabatic index is analyzed. 
The pressure of the star, at the beginning of the collapse, is isotropic but
due to the presence of the shear the pressure becomes more and more
anisotropic.  The black hole is never formed because the apparent horizon
formation condition is never satisfied, in contrast of the previous model where
a black hole is formed.  An observer at infinity sees a
radial point source radiating exponentially until reaches the time of maximum
luminosity and suddenly the star turns off. In contrast of the former model
where the luminosity also increases exponentially, reaching a maximum and after
it decreases until the formation of the black hole.  The effective adiabatic index
is always positive without any discontinuity in contrast of the former model
where there is a discontinuity around the time of maximum luminosity.
The collapse is about three thousand times slower than in the case 
where the energy density is initially homogeneous.
\end{abstract}

\maketitle


\section{Introduction}

One of the most outstanding problems in gravitation theory is the evolution of
a collapsing massive star, after it has exhausted its nuclear fuel.
The problem of constructing physically realistic models for radiating collapsing
stars is one of the aims of the relativistic astrophysics.  However, in order to obtain
realistic models we need to solve complicated systems of nonlinear differential 
equations.   In many cases we can simplify the problem considering some restrictions
in these equations and solve the system analytically.  Such models, although 
simplified, are useful to construct simple exact solutions, which are at least
physically reasonable.  This allows a clearer analysis of the main physical
effects at play, and it can be very useful for checking of numerical procedures.

The majority of the previous works have considered 
only shear-free motion of the fluid \cite{deOliveira85}\cite{Bonnor89}\cite{Wagh01}.  
This simplification allows us to 
obtain exact solutions of the Einstein's equations in some cases but it is 
somewhat
unrealistic.  It is also unrealistic to consider heat flow without viscosity 
but 
if viscosity is introduced, it is desirable to allow shear in the fluid motion,
although we have not included the viscosity in this work.
Thus, it is interesting to study solutions that contains shear,
because it plays a very important role in the study
of gravitational collapse, as shown in \cite{Chan97, Chan98a, Chan98b, Chan00,
Chan01, Nogueira04, Pinheiro08, Pinheiro10} and in \cite{Joshi02}.
More recent studies on this subject with and without shear are also found in
\cite{Misthry08}-\cite{Sharif09}.

In the first paper\cite{Chan97}\cite{Chan98a} we have compared two collapsing
model: a shear-free and a shearing model.  We were interested in studying the 
effect of the shearing motion in the evolution of the collapse.
It was shown
that the pressure of the star, at the beginning of the collapse, is isotropic
but due to the presence of the shear the pressure becomes more and more 
anisotropic.  
The anisotropy in self-gravitating systems has been reviewed and discussed the
causes for its appearance in \cite{Herrera97}.  As shown by
\cite{Chan97}\cite{Chan98a} the simplest cause of the presence of anisotropy in a 
self-gravitating
body is the shearing motion of the fluid, because it appears without an
imposition ad-hoc \cite{Chan93}.

In the second paper \cite{Chan00} we have studied a model of a collapsing 
radiating star consisting of an anisotropic fluid with initial homogeneous energy 
density and shear viscosity undergoing 
radial heat flow with outgoing radiation, but without bulk viscosity.

In the third paper \cite{Chan01} we have analyzed a model of a collapsing 
radiating star consisting of an anisotropic fluid with initial homogeneous energy density 
and bulk viscosity 
undergoing radial heat flow with outgoing radiation, but without shear
viscosity.

The aim of this work is to generalize our previous models
by introducing initial inhomogeneous energy density, but without viscosity. 
 
This work is organized as follows.  In Section 2 we present the Einstein's 
field
equations.  In Section 3 we derive the junction conditions.
In Section 4
we present the proposed solution of the field equations.  In Section 5
we describe the model considered in this work for the initial configuration.
In Section 6 we present the energy conditions for an initial inhomogeneous energy
density fluid.
In Section 7 we show the time evolution of the total mass,
luminosity, the effective adiabatic index and
in Section 8 we summarize the main results obtained in this work.

\section{Field Equations}

We assume a spherically symmetric distribution of fluid undergoing dissipation
in the form of heat flow.  While the dissipative fluid collapses it produces
radiation.  The interior spacetime is described by the most
general spherically symmetric metric, using comoving coordinates,

\begin{equation}
ds^2_{-} = -A^2(r,t)dt^2+B^2(r,t)dr^2 + C^2(r,t)(d\theta^2+\sin^2 \theta d\phi^2).
\label{eq:dsi}
\end{equation}

The exterior spacetime is described by Vaidya's \cite{Vaidya53} metric, which represents
an outgoing radial flux of radiation,

\begin{equation}
ds^2_{+}=-\left[ 1- {\frac{2m(v)}{\bf r}} \right]dv^2-2dvd{\bf r}+{\bf r}^2
(d\theta^2+\sin^2 \theta d\phi^2),
\label{eq:dso}
\end{equation}
where $m(v)$ represents the mass of the system inside the boundary surface
$\Sigma$, function of the retarded time $v$.

We assume the interior energy-momentum tensor is given by

\begin{eqnarray}
G_{\alpha \beta}&=&\kappa T_{\alpha \beta}=\kappa \left[ 
(\mu+p_t)u_{\alpha}u_{\beta}+p_tg_{\alpha \beta}+ 
(p-p_t)X_{\alpha}X_{\beta} +\right. \nonumber \\
& & \left. +q_{\alpha}u_{\beta}+q_{\beta}u_{\alpha} \right],
\label{eq:tab}
\end{eqnarray}
where $\mu$ is the energy density of the fluid,
$p$ is the radial pressure, 
$p_t$ is the tangential pressure,
$q^{\alpha}$ is the radial heat flux,
$X_{\alpha}$ is an unit four-vector along the radial direction, 
$u^{\alpha}$ is the four-velocity, 
which have to satisfy $u^{\alpha}q_{\alpha}=0$, 
$X_{\alpha}X^{\alpha}=1$, $X_{\alpha}u^{\alpha}=0$ and $\kappa=8\pi$ 
(i.e., $c=G=1$).

The shearing tensor $\sigma_{\alpha \beta}$ is defined as
\begin{equation}
\sigma_{\alpha \beta}=u_{(\alpha;\beta)}+ \dot u_{(\alpha} u_{\beta)}-\frac{1}{3} \Theta (g_{\alpha \beta}+u_{\alpha}u_{\beta}),
\label{eq:sab}
\end{equation}
with
\begin{equation}
\dot u_{\alpha}=u_{\alpha;\beta}u^{\beta},
\label{eq:aab}
\end{equation}
\begin{equation}
\Theta=u^{\alpha}_{;\alpha},
\label{eq:theta}
\end{equation}
where $\Theta$ is the expansion scalar, the semicolon denotes a covariant 
derivative and the parentheses in the indices mean symmetrizations.

Since we utilize comoving coordinates we have,

\begin{equation}
u^{\alpha}=A^{-1}\delta^{\alpha}_0,
\label{eq:u}
\end{equation}
and since the heat flux is radial
\begin{equation}
q^{\alpha}=q\delta^{\alpha}_1.
\label{eq:qa}
\end{equation}

Thus the non-zero components of the shearing tensor are given by
\begin{equation}
\sigma_{11}={2 \frac{B^2}{3A}} 
\left( \frac{\dot B}{B} - \frac{\dot C}{C} \right),
\label{eq:sigma11}
\end{equation}
\begin{equation}
\sigma_{22}=-\frac{C^2}{3A}
\left( \frac{\dot B}{B} - \frac{\dot C}{C} \right),
\label{eq:sigma22}
\end{equation}
\begin{equation}
\sigma_{33}=\sigma_{22} \sin^2 \theta.
\label{eq:sigma33}
\end{equation}

A simple calculation shows that
\begin{equation}
\sigma_{\alpha \beta} \sigma^{\alpha \beta}=\frac{2}{3A^2} 
 \left( {\dot B \over B} - {\dot C \over C} \right)^2.
\label{eq:sigma2}
\end{equation}

Thus, if we define the scalar $\sigma$ as
\begin{equation}
\sigma=-{1 \over {3A}} 
 \left( {\dot B \over B} - {\dot C \over C} \right),
\label{eq:sigma}
\end{equation}
we can write that
\begin{equation}
\sigma_{11}=-{2 B^2 \sigma}, 
\label{eq:sigma11a}
\end{equation}
\begin{equation}
\sigma_{22}={C^2 \sigma},
\label{eq:sigma22a}
\end{equation}
\begin{equation}
\sigma_{33}={C^2 \sigma \sin^2 \theta}.
\label{eq:sigma33a}
\end{equation}

Using (\ref{eq:dsi}) and (\ref{eq:theta}), we can write that
\begin{equation}
\Theta={1 \over A}\left({{\dot B \over B} + 2 {\dot C \over C}}\right).
\label{eq:theta1}
\end{equation}

The non-vanishing components of the field equations, using (\ref{eq:dsi}),
(\ref{eq:tab}), (\ref{eq:u}), (\ref{eq:qa}) and 
(\ref{eq:theta}), interior of the boundary
surface $\Sigma$ are

\begin{eqnarray}
G^{-}_{00} &=& -{ \left( A \over B \right) }^2 \left[ 2 {C'' \over C} +
{ \left( C' \over C \right) }^2 - 2{C' \over C}{B' \over B} \right]+ \nonumber \\
& & + {\left( A \over C \right)}^2 + {\dot C \over C}{ \left( {\dot C \over C}
+ 2{\dot B \over B} \right) } = \kappa A^2\mu,
\label{eq:g00}
\end{eqnarray}

\begin{eqnarray}
G^{-}_{11} &=& {C' \over C} { \left( {C' \over C} + 2{A' \over A} \right) }-
{ \left( B \over C \right) }^2 -\nonumber \\
& &- { \left( B \over A \right) }^2
\left[ 2{\ddot C \over C} + { \left( \dot C \over C \right) }^2 -
2{\dot A \over A} {\dot C \over C} \right] \nonumber \\
& &= \kappa B^2 p,
\label{eq:g11}
\end{eqnarray}

\begin{eqnarray}
G^{-}_{22} &=& 
{ \left( {C \over B} \right) }^2 { \left[ {C'' \over C} + {A'' \over A}+
{C' \over C}{A' \over A} - {A' \over A}{B' \over B} - {B' \over B}{C' \over C}
\right] } +\nonumber \\
& & + { \left( C \over A \right) }^2 { \left[ -{\ddot B \over B} -
{\ddot C \over C} - {\dot C \over C}{\dot B \over B} + {\dot A \over A}
{\dot C \over C} + {\dot A \over A}{\dot B \over B} \right] } \nonumber \\
& &= \kappa C^2 p_t,
\label{eq:g22}
\end{eqnarray}

\begin{eqnarray}
G^{-}_{33} &=& {G^{-}_{22} \sin^2 \theta},
\label{eq:g33}
\end{eqnarray}

\begin{eqnarray}
G^{-}_{01} &=& -2{\dot C' \over C} + 2{C' \over C}{\dot B \over B}+
2{A' \over A}{\dot C \over C}= -\kappa A B^2 q.
\label{eq:g01}
\end{eqnarray}

The dot and the prime stand for differentiation with respect to $t$ and $r$,
respectively.

\section{Junction Conditions}

We consider a spherical surface with its motion described by a time-like
three-space $\Sigma$, which divides spacetimes into interior and exterior
manifolds.  For the junction conditions we follow the approach given by
\cite{Israel66a}\cite{Israel66b}.  Hence we have to demand

\begin{equation}
(ds^2_{-})_{\Sigma}=(ds^2_{+})_{\Sigma},
\label{eq:dsidso}
\end{equation}

\begin{equation}
K^{-}_{ij}=K^{+}_{ij},
\label{eq:kijikijo}
\end{equation}
where $K^{\pm}_{ij}$ is the extrinsic curvature to $\Sigma$, given by

\begin{equation}
K^{\pm}_{ij}=-n^{\pm}_{\alpha}
{{\partial^2x^{\alpha}_{\pm}} \over {\partial \xi^i \partial \xi^j}}
-n^{\pm}_{\alpha}\Gamma^{\alpha}_{\beta \gamma}
{{\partial x^{\beta}_{\pm}}  \over {\partial \xi^i}}
{{\partial x^{\gamma}_{\pm}} \over {\partial \xi^j}},
\label{eq:kij}
\end{equation}
and where $\Gamma^{\alpha}_{\beta \gamma}$ are the Christoffel symbols, 
$n^{\pm}_{\alpha}$ the unit normal vectors to $\Sigma$, $x^{\alpha}$ are
the coordinates of interior and exterior spacetimes and $\xi^i$ are the 
coordinates that define the surface $\Sigma$.

From the junction conditions (\ref{eq:dsidso}) and (\ref{eq:kijikijo}) we obtain the following
results (see more details in \cite{Chan97}-\cite{Chan01})

\begin{equation}
m=\left\{ {C \over 2}\left[ 1 + {\left( \dot C \over A \right)}^2 -
{\left( C' \over B \right)}^2 \right] \right\}_{\Sigma},
\label{eq:ms}
\end{equation}
which is the total energy entrapped inside the surface $\Sigma$ \cite{Cahill70},

\begin{equation}
p=(qB)_{\Sigma},
\label{eq:pqbs}
\end{equation}
(This result is analogous to the one obtained by \cite{Chan97}\cite{Chan98a}\cite{Chan00}\cite{Chan01} 
for a shearing fluid motion.)

\begin{equation}
L_{\infty}={\kappa \over 2}\left[ p C^2 { \left( {C' \over B} + 
{\dot C \over A} \right) }^2 \right]_{\Sigma},
\label{eq:lsp}
\end{equation}
which is the total luminosity for an observer at rest at infinity
and

\begin{equation}
1+z_{\Sigma}=
{ \left( {C' \over B} + {\dot C \over A} \right)^{-1} }_{\Sigma}.
\label{eq:zs}
\end{equation}
which is the boundary redshift $z_{\Sigma}$. 
The boundary redshift can be used to determine the time of formation of
the horizon.

\section{Solution of the Field Equations}

As in \cite{Chan97}\cite{Chan98a}\cite{Chan00}\cite{Chan01} we are proposing solutions of the field 
equations (\ref{eq:g00})-(\ref{eq:g01})
with the form

\begin{equation}
A(r,t)=A_0(r),
\label{eq:art0}
\end{equation}

\begin{equation}
B(r,t)=B_0(r),
\label{eq:brt0}
\end{equation}

\begin{equation}
C(r,t)=rB_0(r)f(t),
\label{eq:crt0}
\end{equation}
where $A_0(r)$ and $B_0(r)$ are solutions of a static perfect fluid having
$\mu_0$ as the energy density and $p_0$ as the isotropic pressure.  

We have chosen this separation of variables in the metric functions, in order
to have the following properties: when $f(t) \rightarrow 1$ the metric functions 
represent the static solution of the initial star configuration and the
collapse takes place when $f(t) \rightarrow 0$.
We stress here that, following the junction condition, equation (\ref{eq:dsidso}), 
the function
$C(r_{\Sigma},t)$ represents the luminosity radius of the body as seen by an 
exterior observer.

Thus, the expansion scalar (\ref{eq:theta1}) can be written as
\begin{equation}
\Theta={2 \over {A_0}} \left( {{\dot f \over f}} \right),
\label{eq:thetaa}
\end{equation}
and the shear scalar (\ref{eq:sigma}) can be written as
\begin{equation}
\sigma={1 \over {3A_0}} \left({\dot f \over f}\right).
\label{eq:sigmaa}
\end{equation}

Now the equations (\ref{eq:g00})-(\ref{eq:g01}) can be written as

\begin{equation}
\kappa \mu = \kappa {\mu_0}+ {1 \over {A^2_0}}\left( \dot f \over f \right)^2+
{1 \over {r^2B^2_0}}\left( {1 \over f^2} - 1 \right)
\label{eq:mu}
\end{equation}

\begin{eqnarray}
\kappa p&=&\kappa{p_0} - {1 \over {A^2_0}}\left[ 2{\ddot f \over f}+
{\left( \dot f \over f \right)}^2 \right]-
{1 \over {r^2B^2_0}}\left( {1 \over f^2} - {1} \right),
\label{eq:p}
\end{eqnarray}

\begin{equation}
\kappa p_t=\kappa {p_0} - {1 \over {A_0^2}} \left( { \ddot f \over f} \right)
\label{eq:pt}
\end{equation}

\begin{equation}
\kappa q = {2 \over {A_0B^2_0}} \left[ \left( {\dot f \over f} \right)
\left( {B'_0 \over B_0} + {1 \over r} - {A'_0 \over A_0} \right) \right],
\label{eq:q}
\end{equation}
where

\begin{equation}
\kappa \mu_0 = -{1 \over {B^2_0}}\left[ 2{B''_0 \over B_0} - 
{\left( B'_0 \over B_0 \right)}^2 + {4 \over r}{B'_0 \over B_0} \right],
\label{eq:mu0}
\end{equation}

\begin{equation}
\kappa p_0 = {1 \over {B^2_0}}\left[ {\left( B'_0 \over B_0 \right)}^2 + 
{2 \over r}{B'_0 \over B_0} + 2{A'_0 \over A_0}{B'_0 \over B_0} +
{2 \over r}{A'_0 \over A_0} \right].
\label{eq:p0}
\end{equation}

We can see from equations (\ref{eq:mu})-(\ref{eq:q}) that when the function
$f(t)=1$ we obtain the static perfect fluid configuration.

Substituting equations (\ref{eq:p}), (\ref{eq:q}) and 
(\ref{eq:thetaa})-(\ref{eq:sigmaa}) into 
(\ref{eq:pqbs}), assuming also that $p_0(r_{\Sigma})=0$,
we obtain a second order differential equation in $f(t)$,

\begin{equation}
2{\ddot f \over f} + {\left( \dot f \over f \right)}^2
+a\left( {\dot f \over f} \right) +
b\left( {1 \over f^2} - {1} \right)=0,
\label{eq:pqbsigma}
\end{equation}
where

\begin{equation}
a = \left[ 2{\left(A_0 \over B_0 \right)}
\left( {B'_0 \over B_0} + {1 \over r} - {A'_0 \over A_0} \right)
\right]_{\Sigma},
\label{eq:as}
\end{equation}
and
\begin{equation}
b = \left( A^2_0 \over {r^2B^2_0} \right)_{\Sigma}.
\label{eq:bs}
\end{equation}

Now, we can write equation (\ref{eq:pqbsigma}) in the following way

\begin{equation}
2f\ddot f + {\dot f}^2 + a f \dot f + b(1-f^2)=0,
\label{eq:ft}
\end{equation}

\begin{figure}
\vspace{.2in}
\centerline{\psfig{figure=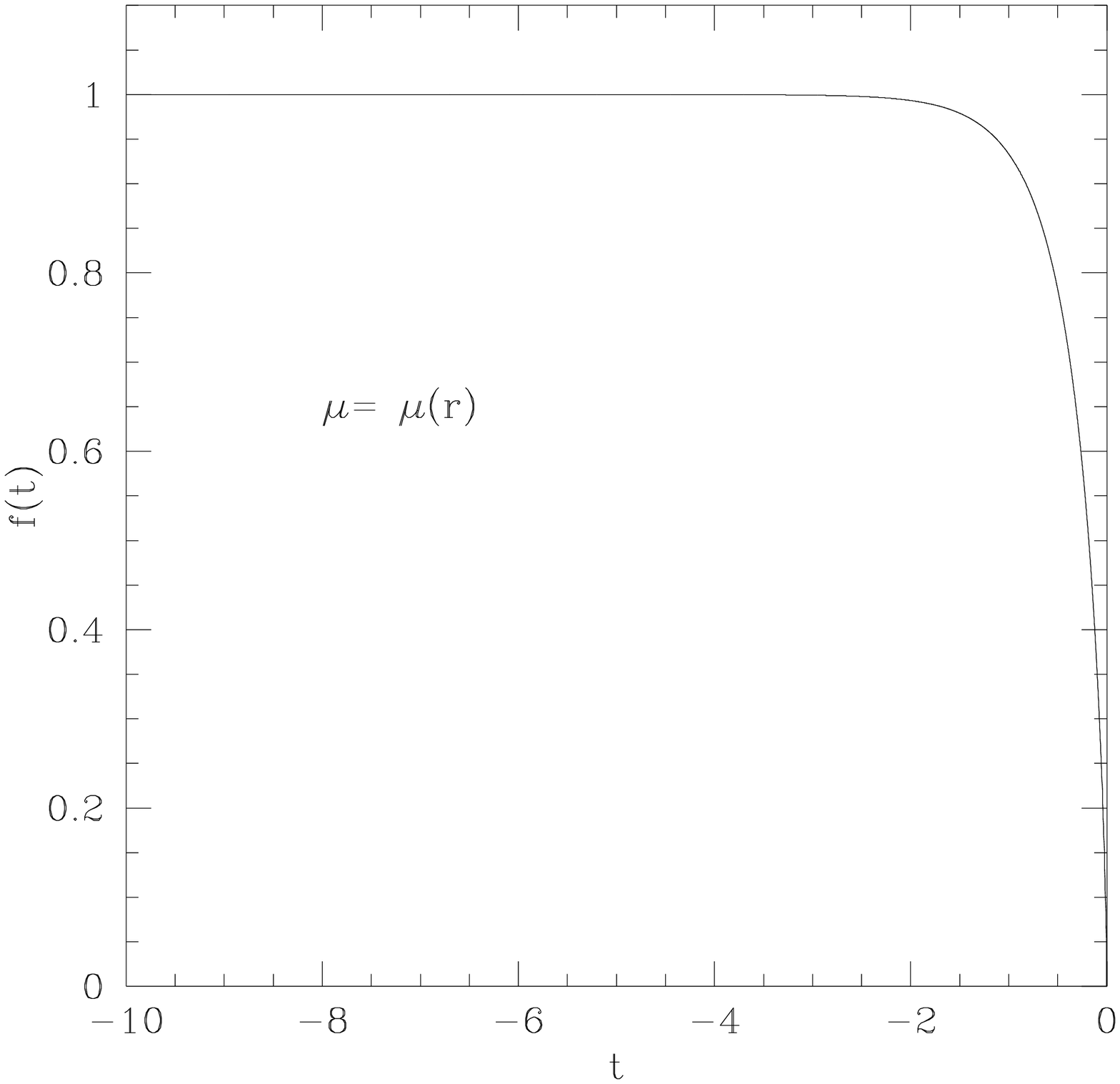,width=3.3truein,height=3.0truein}\hskip
.25in \psfig{figure=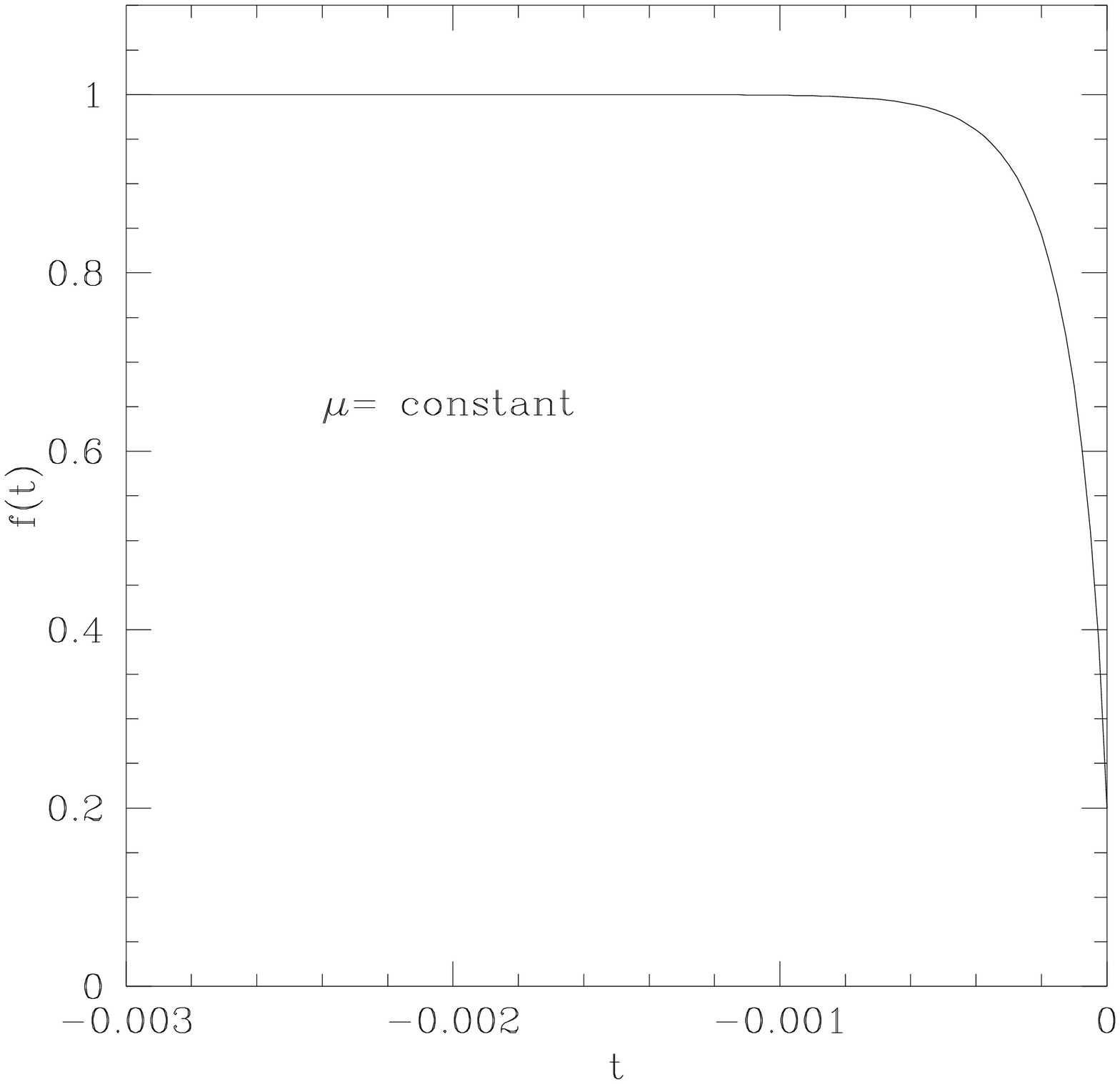,width=3.3truein,height=3.0truein}
\hskip .5in} \caption{Time behavior of the function $f(t)$ for the models with
homogeneous and initial inhomogeneous density.  The time is in units of second and 
$f(t)$ is dimensionless.  The collapse for the initial homogeneous energy density model ($\mu = constant$)
is 3000 times faster than that one for the initial inhomogeneous energy density model ($\mu=\mu(r)$).}
\label{foft}
\end{figure}

This equation is identical to the one obtained in 
\cite{Chan97}\cite{Chan98a}\cite{Chan00}\cite{Chan01}.  Thus, as before
it has to be solved numerically (figure \ref{foft}), assuming that at 
$t\rightarrow-\infty$ represents the static configuration with
$\dot f (t\rightarrow-\infty)\rightarrow0$ and 
$f(t\rightarrow-\infty)\rightarrow1$.  We also assume that $f(t\rightarrow 0)
\rightarrow 0$.
This means that the luminosity radius $C(r_{\Sigma},t)$ has the value 
$r_{\Sigma}B_0(r_{\Sigma})$ at the beginning of the collapse and vanishing at 
the end of the evolution.

\section{Model of the Initial Configuration}

We consider that the system at the beginning of the collapse has a static
configuration of a perfect fluid satisfying \cite{Goldman78}

\begin{equation}
A_0^2=d_0\left[ \frac{1+(\alpha+\beta)g(r)}{1+(\alpha-\beta)g(r)} \right]^{1/\beta},
\label{eq:a0}
\end{equation}

\begin{equation}
B_0^2=c_0\frac{g(r)^2}{a_0^2(1+2\alpha g(r) + 2g(r)^2)} 
\left[ \frac{1+(\alpha-\beta)g(r)}{1+(\alpha+\beta)g(r)} \right]^{(\alpha+1)/\beta},
\label{eq:b0}
\end{equation}
where

\begin{equation}
g(r)=\frac{a_0}{1-r^2},
\label{eq:gr}
\end{equation}
\begin{equation}
\beta=\sqrt{\alpha^2 - 2},
\label{eq:beta}
\end{equation}
\begin{equation}
d_0=\left( \frac{2r_{\Sigma}-m_0}{2r_{\Sigma}+m_0} \right)^2 
\left[ \frac{1+(\alpha-\beta)g(r_{\Sigma})}{1+(\alpha+\beta)g(r_{\Sigma})} \right]^{1/\beta},
\label{eq:dd}
\end{equation}
\begin{equation}
c_0=\frac{a_0^2}{g(r_{\Sigma})}\left( 1+ \frac{m_0}{2r_{\Sigma}} \right)^4 
\left[ 1+2\alpha g(r_{\Sigma}) + 2g(r_{\Sigma})^2 \right]
\left[ \frac{1+(\alpha+\beta)g(r_{\Sigma})}{1+(\alpha-\beta)g(r_{\Sigma})} \right]^{(\alpha+1)/\beta},
\label{eq:cc}
\end{equation}
\begin{equation}
a_0=\frac{2r_{\Sigma}}{m_0} (1-r_{\Sigma}^2),
\label{eq:aa}
\end{equation}
\begin{equation}
\alpha=\frac{r_{\Sigma}^3}{m_0 (1-r_{\Sigma}^2)}-\frac{m_0}{4r_{\Sigma}(1-r_{\Sigma}^2)}-\frac{2r_{\Sigma}}{m_0},
\label{eq:alpha}
\end{equation}
and where $r_{\Sigma}$ is the initial coordinate radius of the star in comoving coordinates
and $m_0$ is the initial mass of the system.
Thus the static initial inhomogeneous energy density and static pressure are given by

\begin{eqnarray}
\kappa \mu_0 &=&\frac{1}{B_0^2\left[ (1-r^2)^2 + 2\alpha a_0 (1-r^2) + 2a_0^2 \right]^2} 
\left[ -4(2r^2-3a_0+1+4\alpha a_0)(1-r^2)^2- \right. \nonumber \\
& & \left. 8(1-r^2)^3 -8(\alpha a_0 -3 r^2a_0 -3\alpha a_0^2 -r^2 \alpha a_0 + 2a_0^2)(1-r^2)- \right. \nonumber \\
& & \left. 4a_0^2(2-6a_0-4\alpha r^2-5r^2) \right],
\label{eq:mu0a}
\end{eqnarray}

\begin{eqnarray}
\kappa p_0 &=&\frac{1}{B_0^2\left[ (1-r^2)^2 + 2\alpha a_0 (1-r^2) + 2a_0^2 \right]^2} \times \nonumber \\
& & \left[ 4(1-r^2)^3+(4r^2+8\alpha a_0)(1-r^2)^2+8a_0^2(1-r^2)-4a_0^2r^2 \right].
\label{eq:p0a}
\end{eqnarray}
This solution shows that the pressure and the energy density are finite and positive
everywhere inside the star, the pressure and energy density decreases toward the
boundary and that the speed of sound is smaller than the speed of the light $c$
everywhere.

We consider the initial configuration as due to an helium core of a 
presupernova with $m_0=6M_{\odot}$, 
with $r_{\Sigma}=2.1232\times10^5$ km which corresponds
approximately the same geometrical radius
$r_{\Sigma}B_0(r_{\Sigma}$) = $2.1218\times10^5$ km \cite{Woosley88}.  

With these 
values we can solve numerically the differential equation (\ref{eq:ft}).  
Since the equation (\ref{eq:a0}) gives us positive or negative values for
$A_0$, we have assumed that $A_0 < 0$.
Thus, we can see from (\ref{eq:q}), using (\ref{eq:a0})-(\ref{eq:cc}) and this
initial configuration, that 
$[(B'_0/B_0+1/r-A'_0/A_0)/A_0]_{\Sigma} < 0$, $(B'_0/B_0+1/r)_{\Sigma} > 0$,
and by the fact that $q_{\Sigma} > 0$ then we conclude that $\dot f < 0$.

In the figure 1 we can note that the collapse for the initial homogeneous 
energy density model ($\mu = constant$) is 3000 times faster than that one 
for the initial inhomogeneous energy density model ($\mu=\mu(r)$).
This fact may be due to two reasons. Firstly, the geometrical radius of the
both models are not the same. Secondly, the metric functions $A_0(r)$ and 
$B_0(r)$ are also different, in such way that in the present work we have
an initial monotonically decreasing energy density profile.  This issue is
also discussed in previous works \cite{Herrera98}\cite{Herrera99} in the
context of the Tolman mass.  These works have shown that a negative energy
density gradient increases the Tolman mass of the system, thus decreasing
the total collapse time. This result seems to agree with our present work. 
Besides, since the initial inhomogeneous model collapses slower is also
related to the fact that in the present model the system ejects all its mass
(this result will be shown later below) 
reducing drastically the gravitationally force term, unlike the initially 
homogeneous case where only about 33\% of the mass is radiated.

\begin{figure}
\vspace{.2in}
\centerline{\psfig{figure=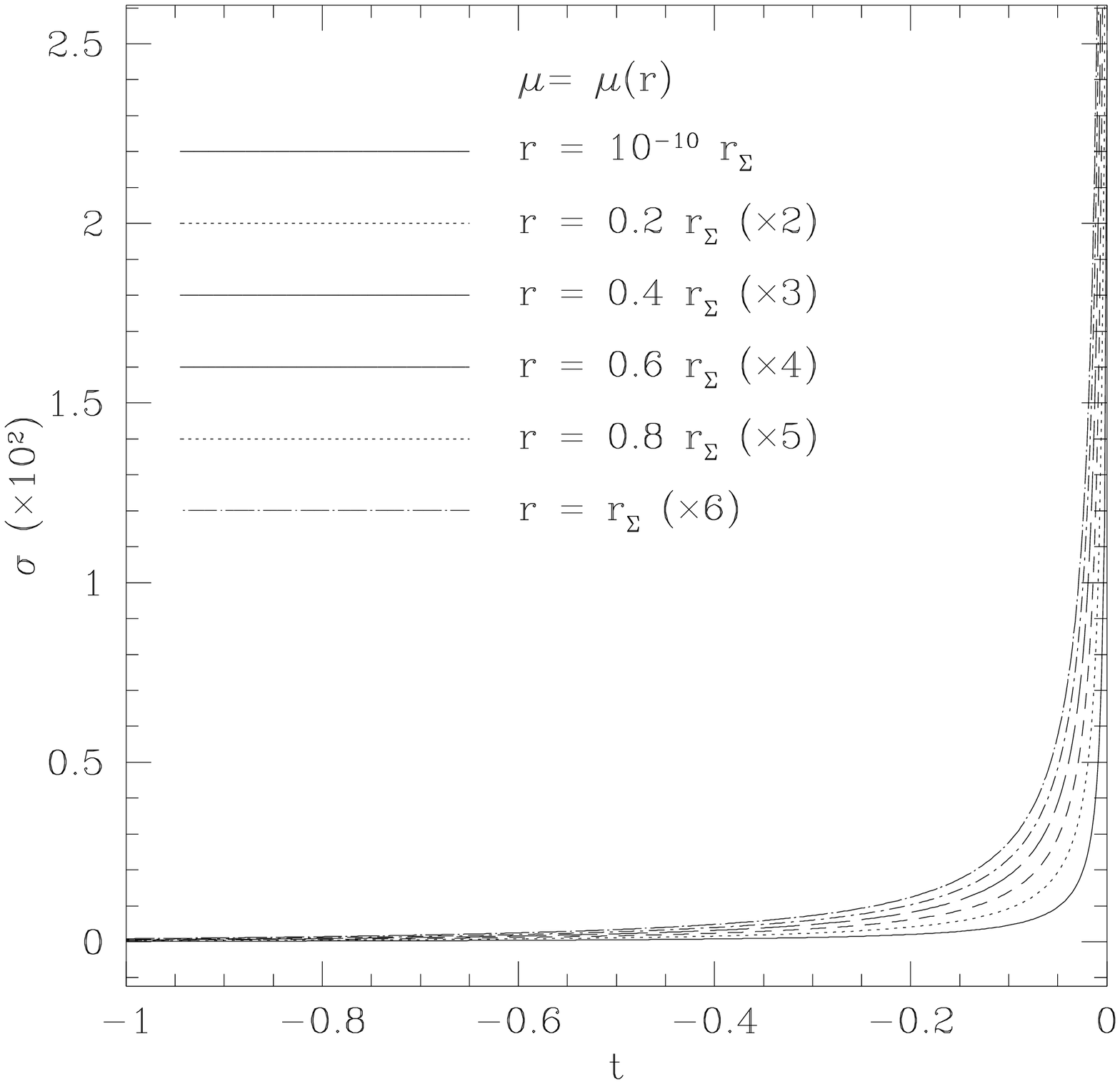,width=3.3truein,height=3.0truein}\hskip
.25in \psfig{figure=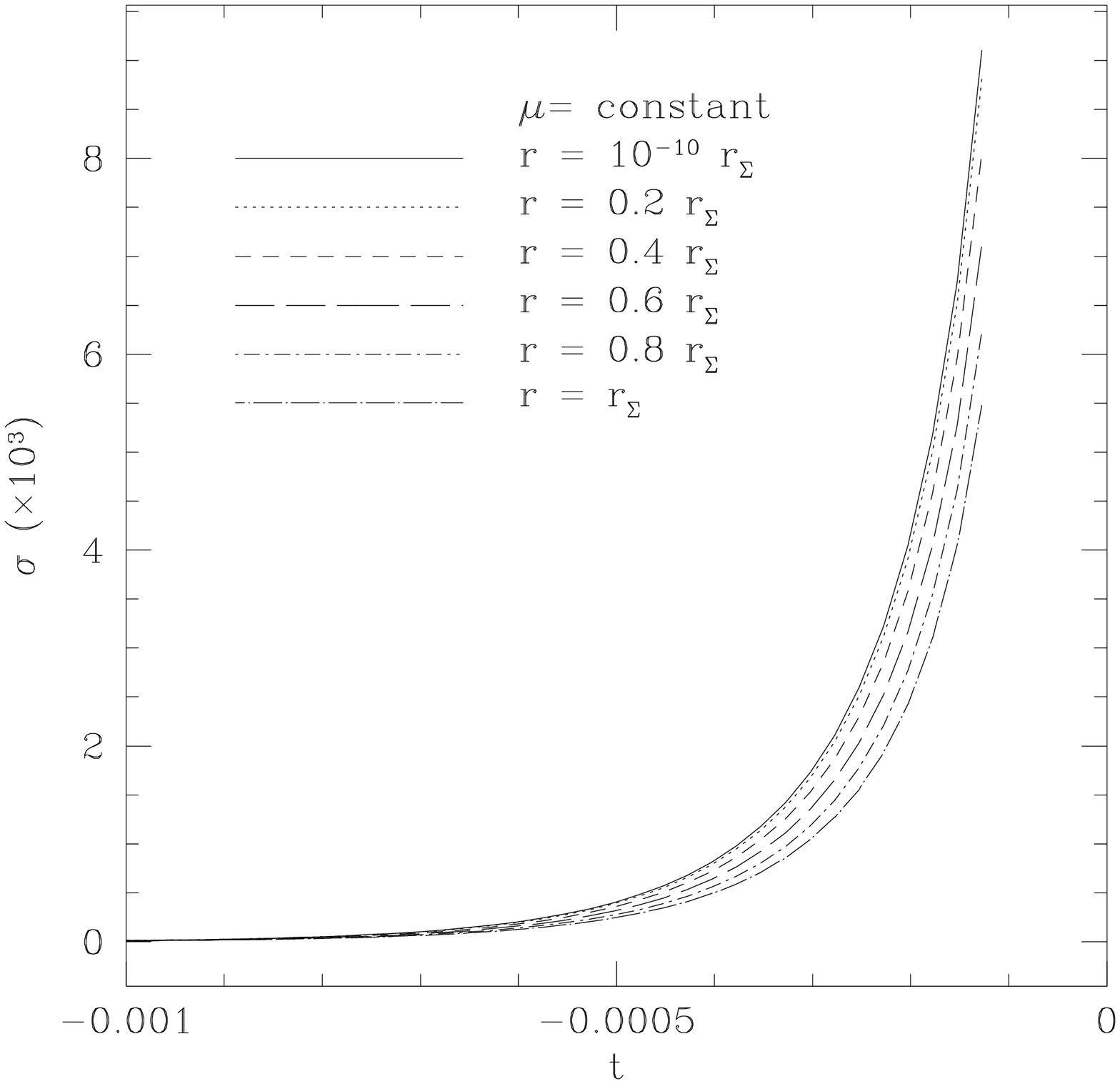,width=3.3truein,height=3.0truein}
\hskip .5in} \caption{Time evolution of the shear scalar at different radii
for the initial inhomogeneous and homogeneous energy density model.  The radii
$r$, $r_{\Sigma}$ and the time are in units of seconds and the shear scalar
 is in units of sec$^{-1}$. } 
\label{sigma}
\end{figure}

In figure \ref{sigma} we show the time evolution of the shear scalar, for
different radii, for the two models (initial inhomogeneous and homogeneous 
energy density).

In order to determine
the time of formation of the horizon $f_{\rm bh}$, we use the equations 
(\ref{eq:zs}) and (\ref{eq:art0})-(\ref{eq:crt0}) and write

\begin{equation}
{\dot f_{\rm bh} \over f_{\rm bh}} = -\left[ {A_0 \over B_0}
\left( {B'_0 \over B_0} + {1 \over r} \right) \right]_{\Sigma}
\approx +1.4140.
\label{eq:fbh1}
\end{equation}

\begin{figure}
\vspace{.2in}
\centerline{\psfig{figure=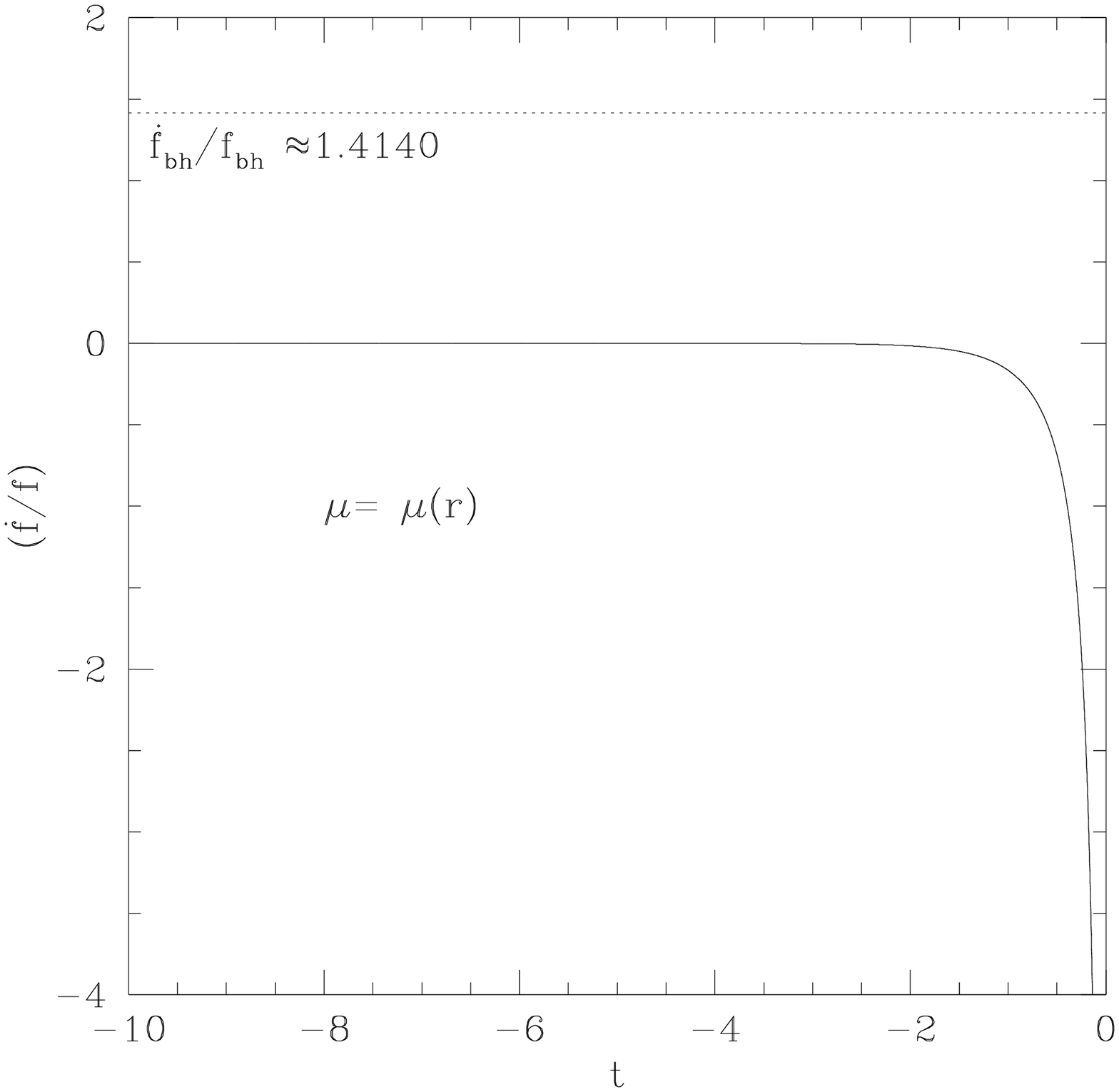,width=3.3truein,height=3.0truein}\hskip
.25in \psfig{figure=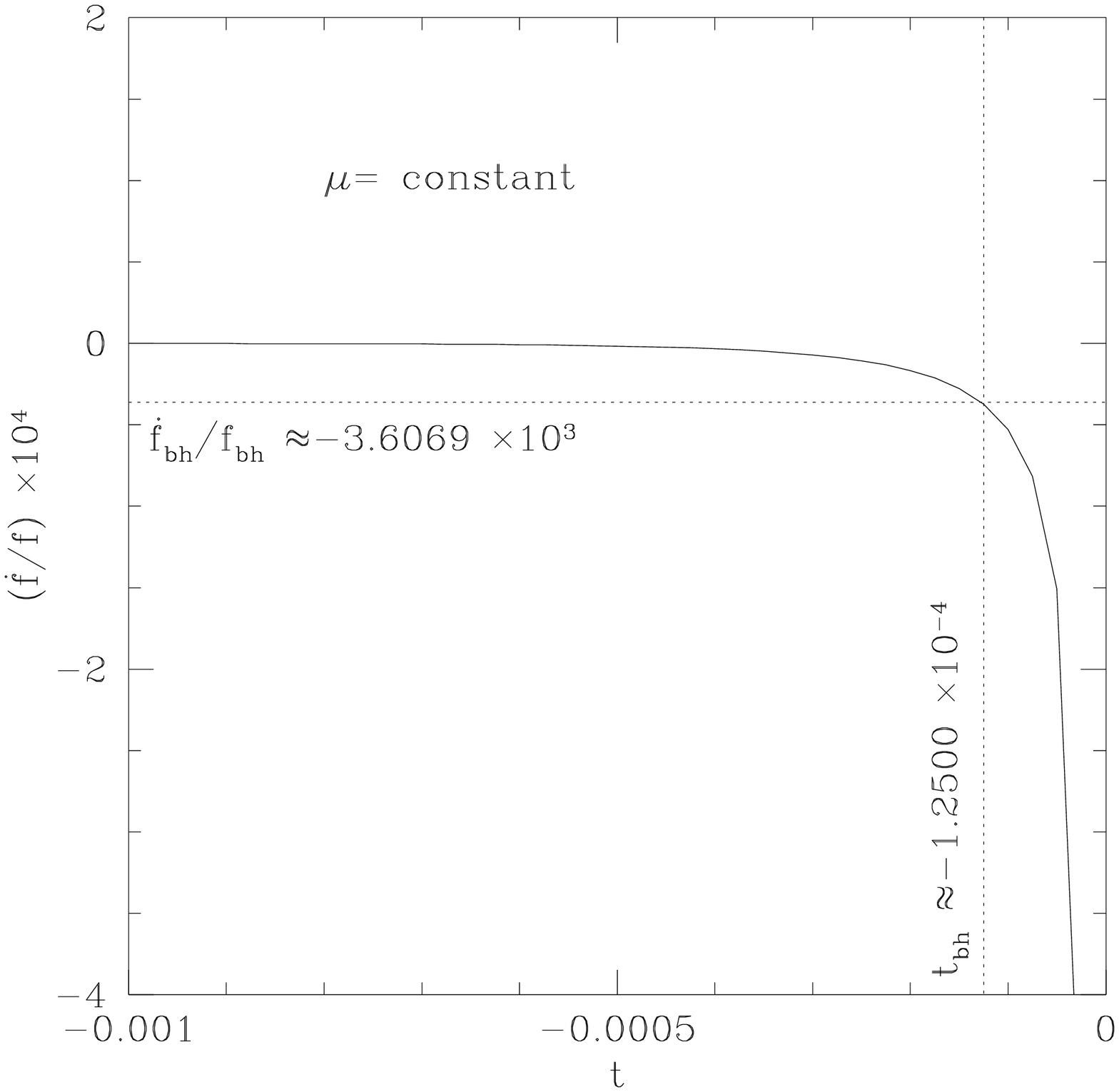,width=3.3truein,height=3.0truein}
\hskip .5in} \caption{The function $\dot f/f$ as a function of the time
for the model with initial homogeneous and inhomogeneous energy density.  
The time is in units of second.}
\label{bh}
\end{figure}

Using the numerical solution of $f(t)$ and equation (\ref{eq:fbh1}),
we can see from figure \ref{bh} that the horizon is never formed, because
the function $\dot f/f$ never reaches the value $+1.4140$.
At the first sight this fact could be interpreted as the formation of a
naked singularity.  However, this is not the case as we will see below in
the calculation of the total energy entrapped inside the hypersurface
$\Sigma$.

\begin{figure}
\vspace{.2in}
\centerline{\psfig{figure=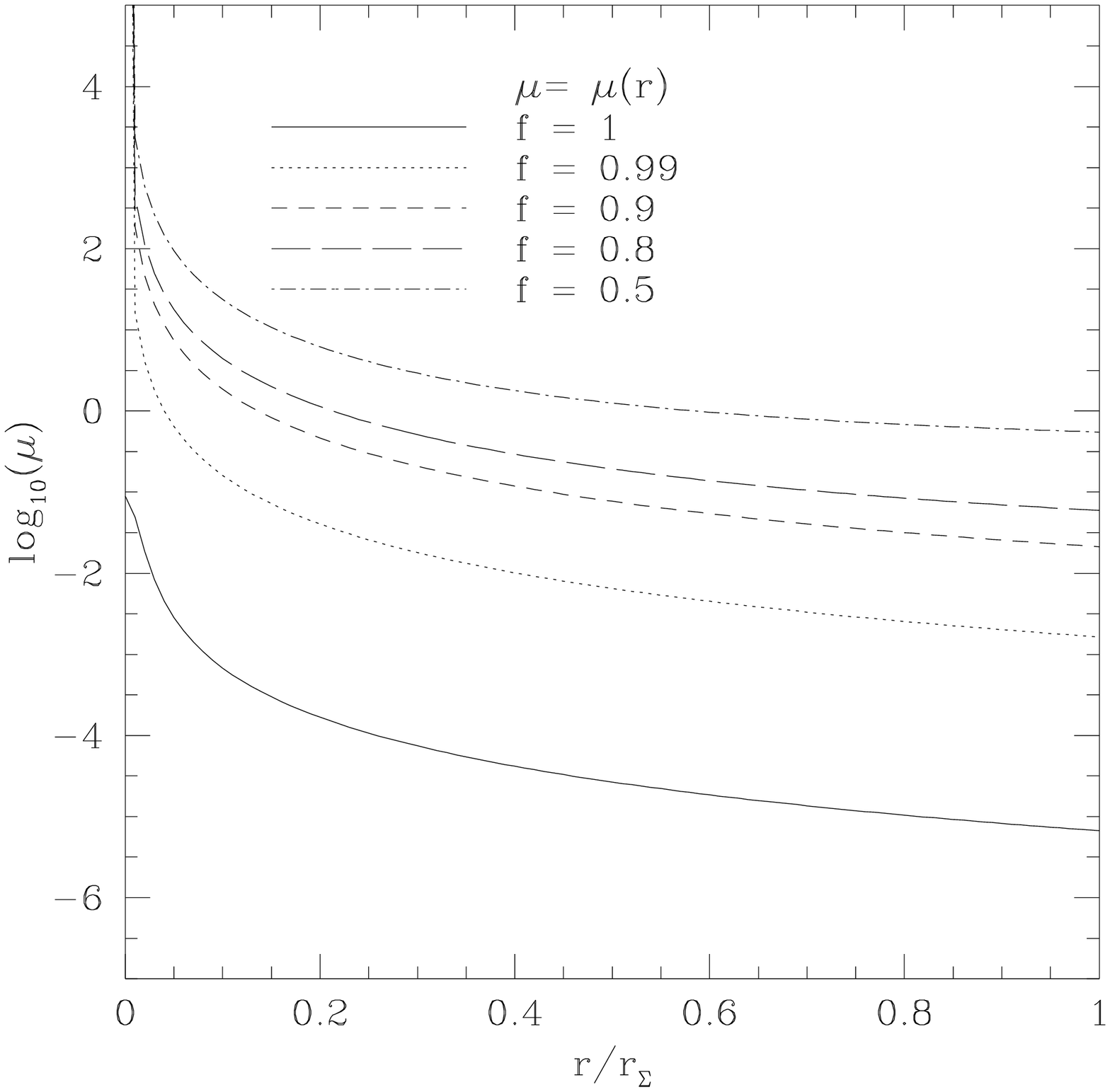,width=3.3truein,height=3.0truein}\hskip
.25in \psfig{figure=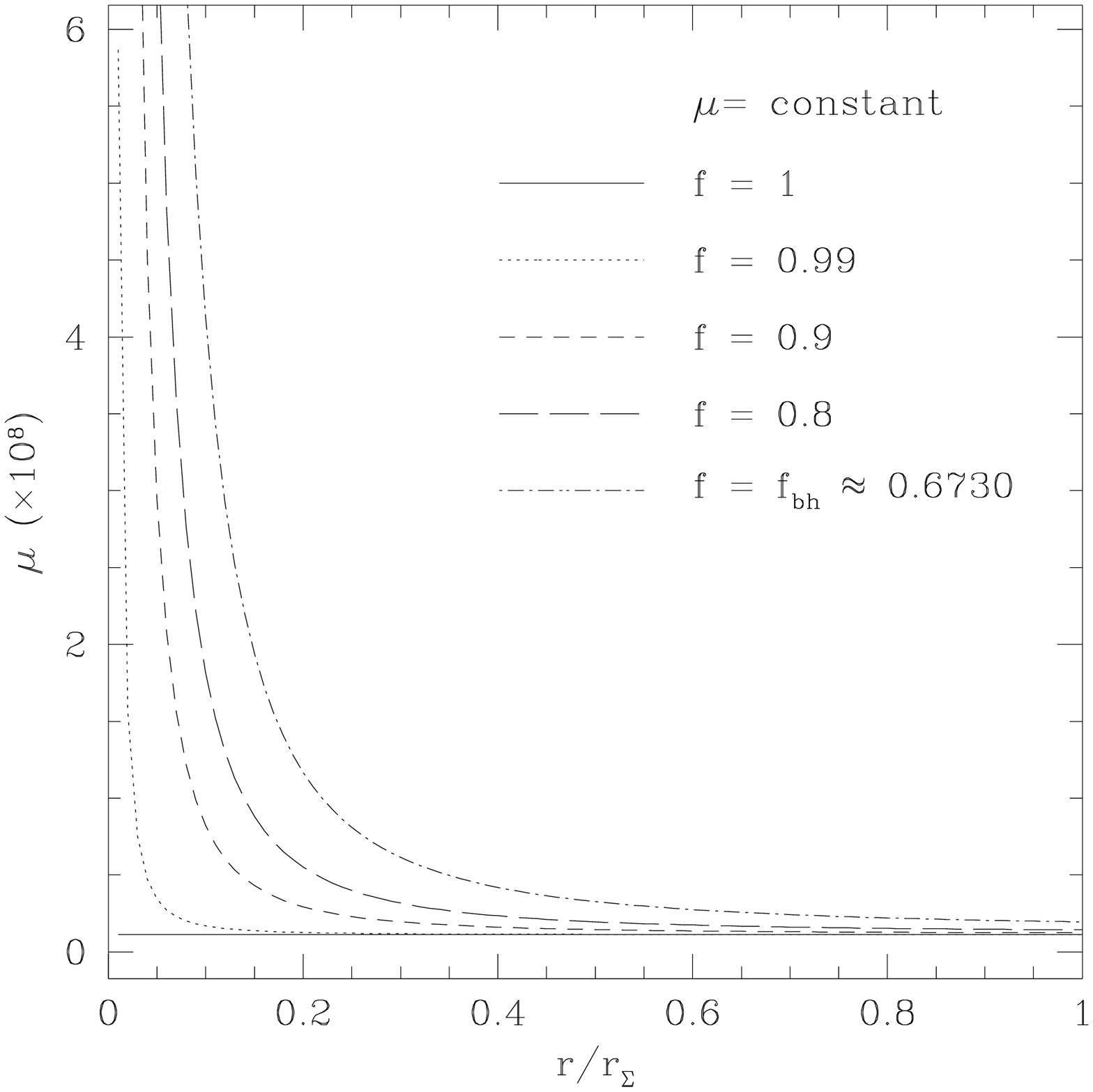,width=3.3truein,height=3.0truein}
\hskip .5in} \caption{Density profiles for the model with initial homogeneous and
initial inhomogeneous energy density.  The radii $r$ and $r_{\Sigma}$ are in units of 
seconds and the density is in units of sec$^{-2}$.}
\label{mu}
\end{figure}

\begin{figure}
\vspace{.2in}
\centerline{\psfig{figure=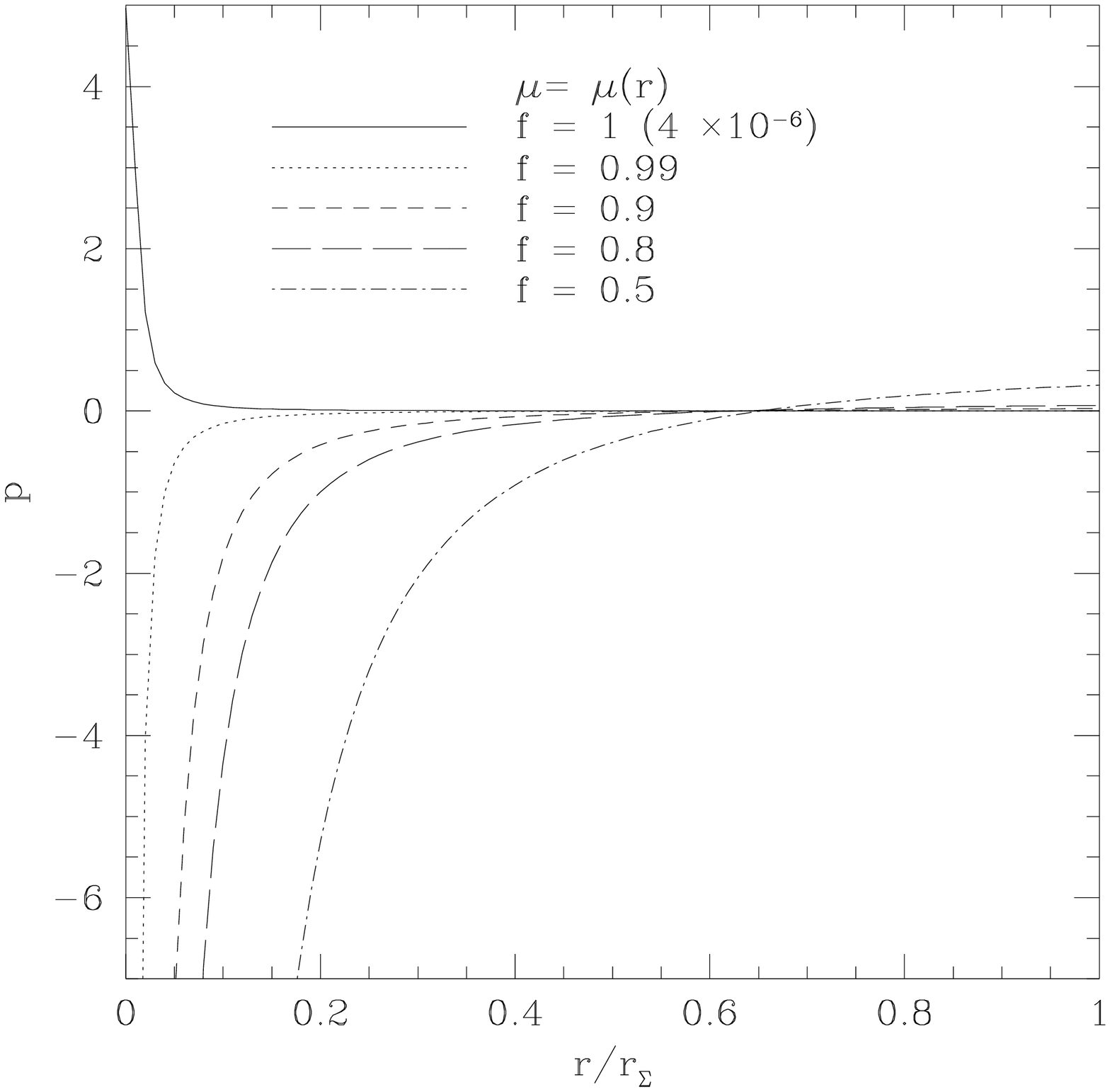,width=3.3truein,height=3.0truein}\hskip
.25in \psfig{figure=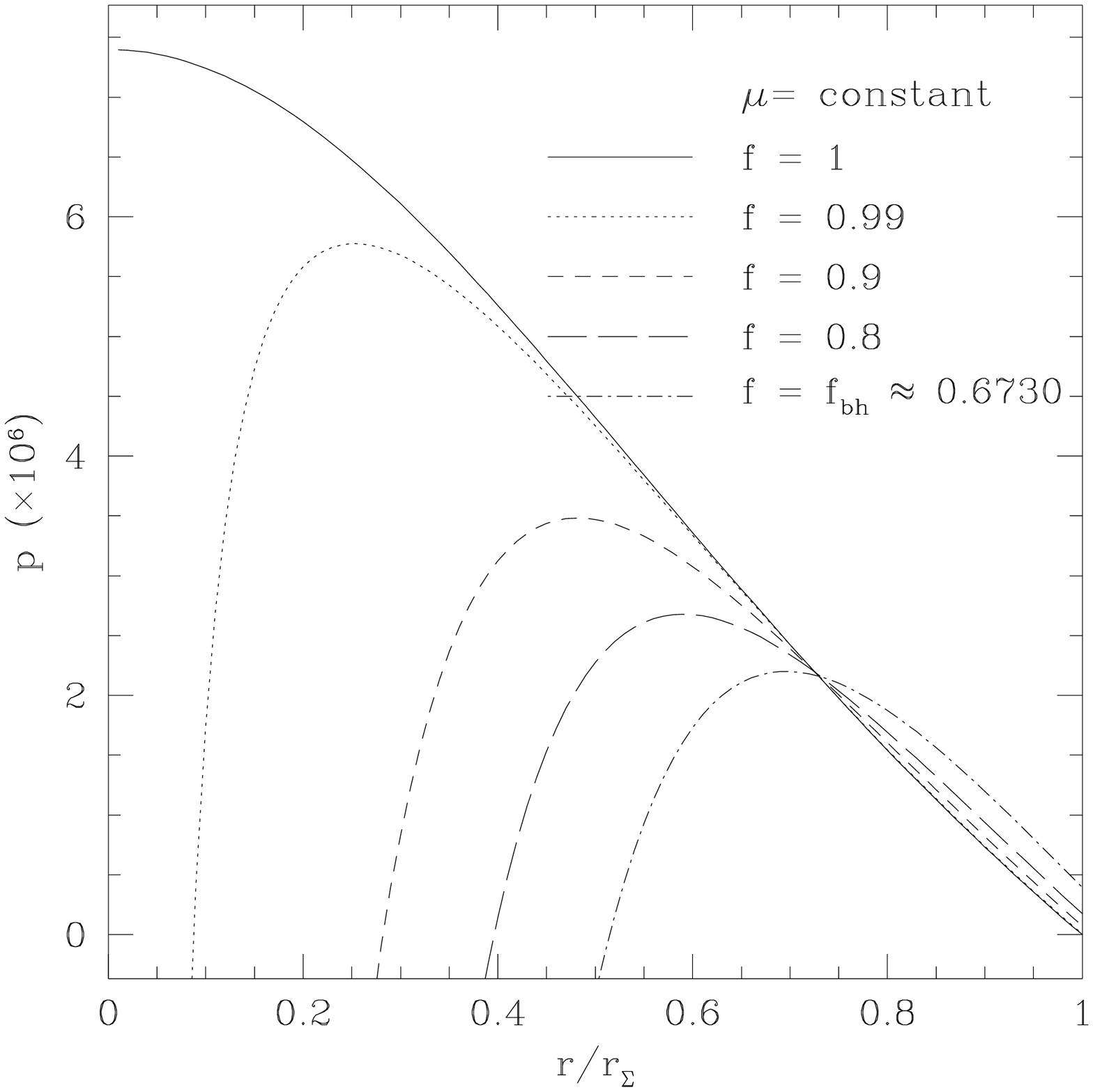,width=3.3truein,height=3.0truein}
\hskip .5in} \caption{Radial pressure profiles for the model with initial homogeneous and
initial inhomogeneous energy density.  The radii $r$ and $r_{\Sigma}$ are in units of 
seconds and the radial pressure is in units of sec$^{-2}$.}
\label{pr}
\end{figure}

\begin{figure}
\vspace{.2in}
\centerline{\psfig{figure=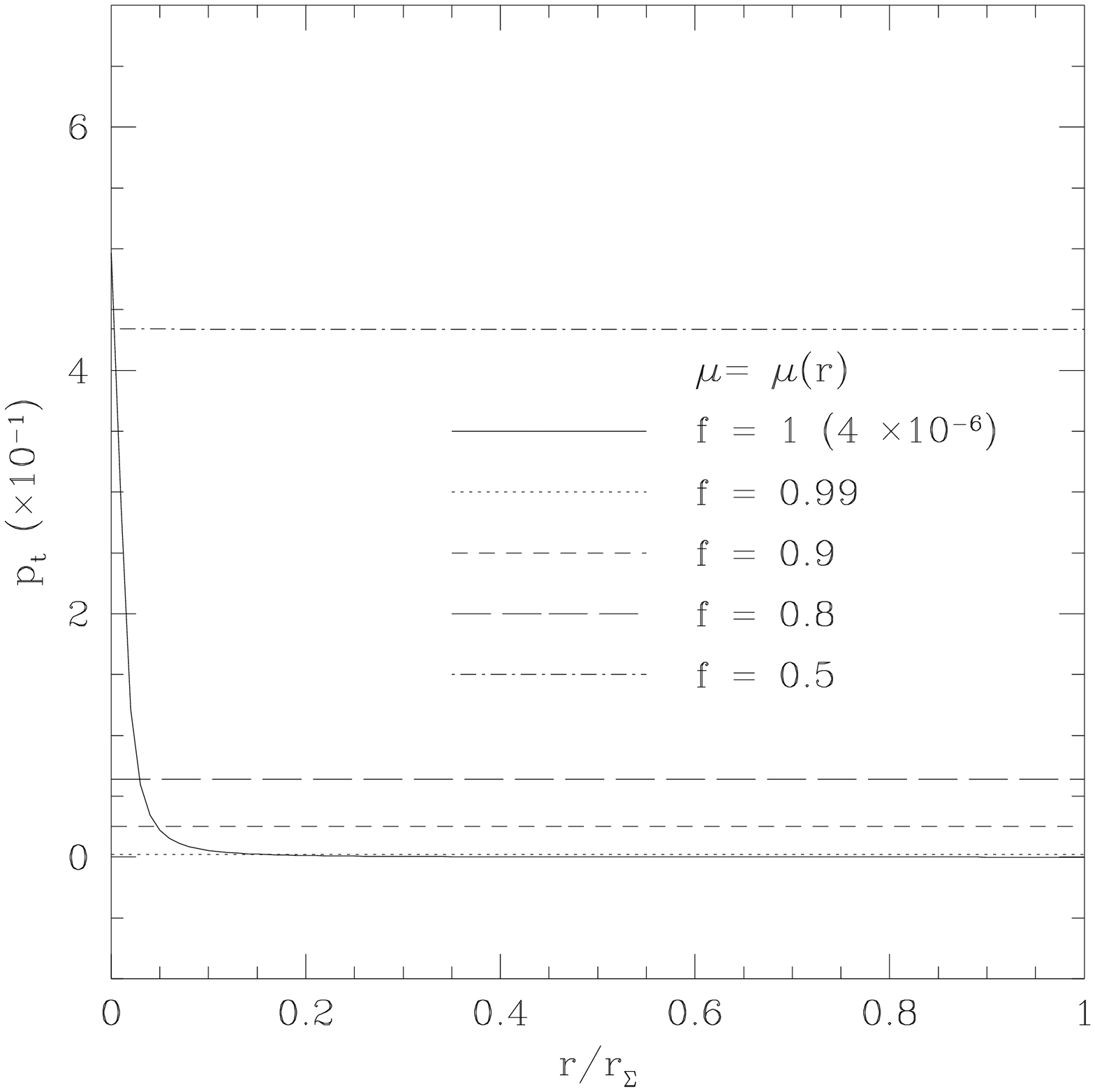,width=3.3truein,height=3.0truein}\hskip
.25in \psfig{figure=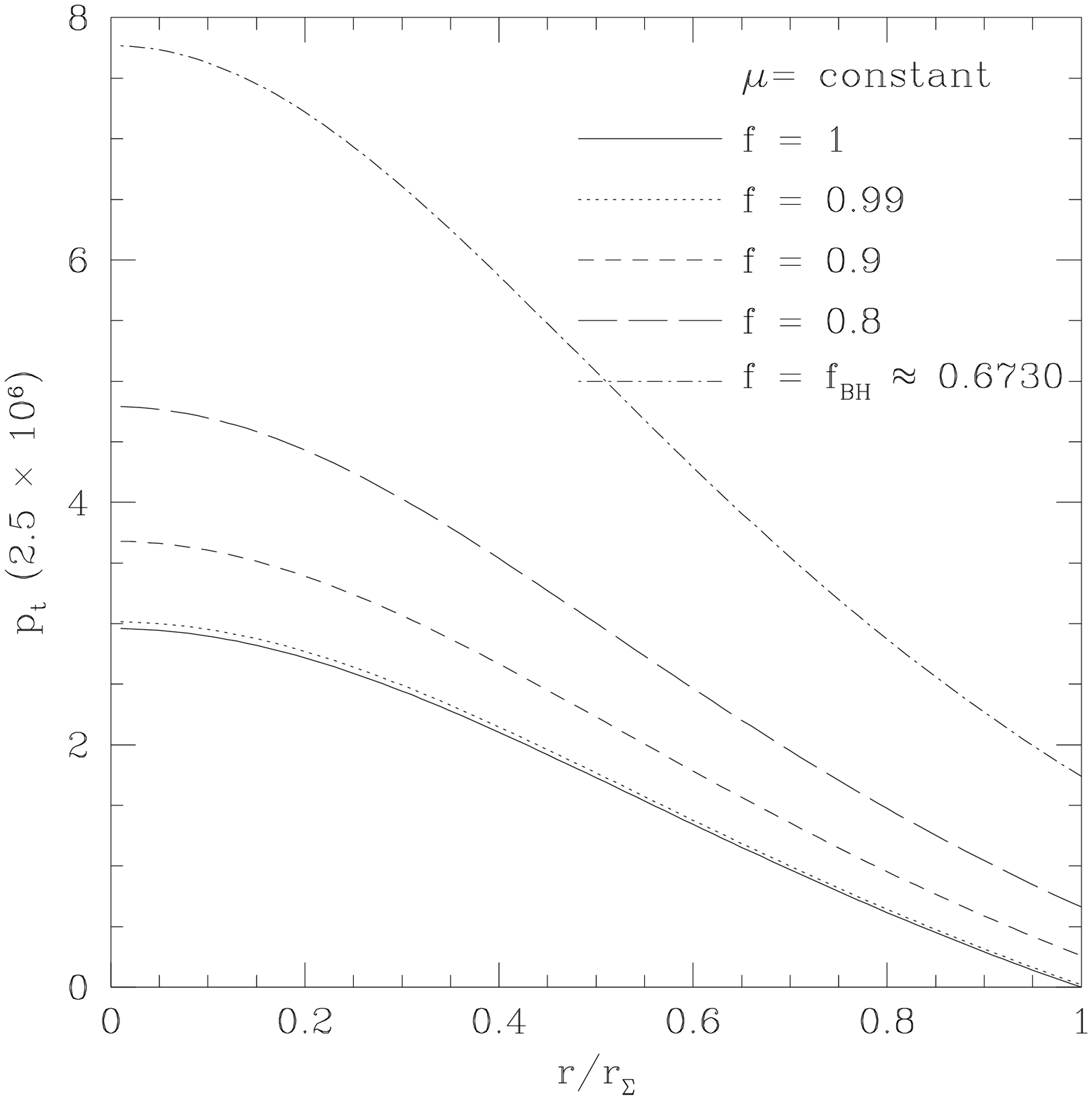,width=3.3truein,height=3.0truein}
\hskip .5in} \caption{Tangential pressure profiles for the model with initial homogeneous and
initial inhomogeneous energy density.  The radii $r$ and $r_{\Sigma}$ are in units of 
seconds and the tangential pressure is in units of sec$^{-2}$.}
\label{pt}
\end{figure}

\begin{figure}
\vspace{.2in}
\centerline{\psfig{figure=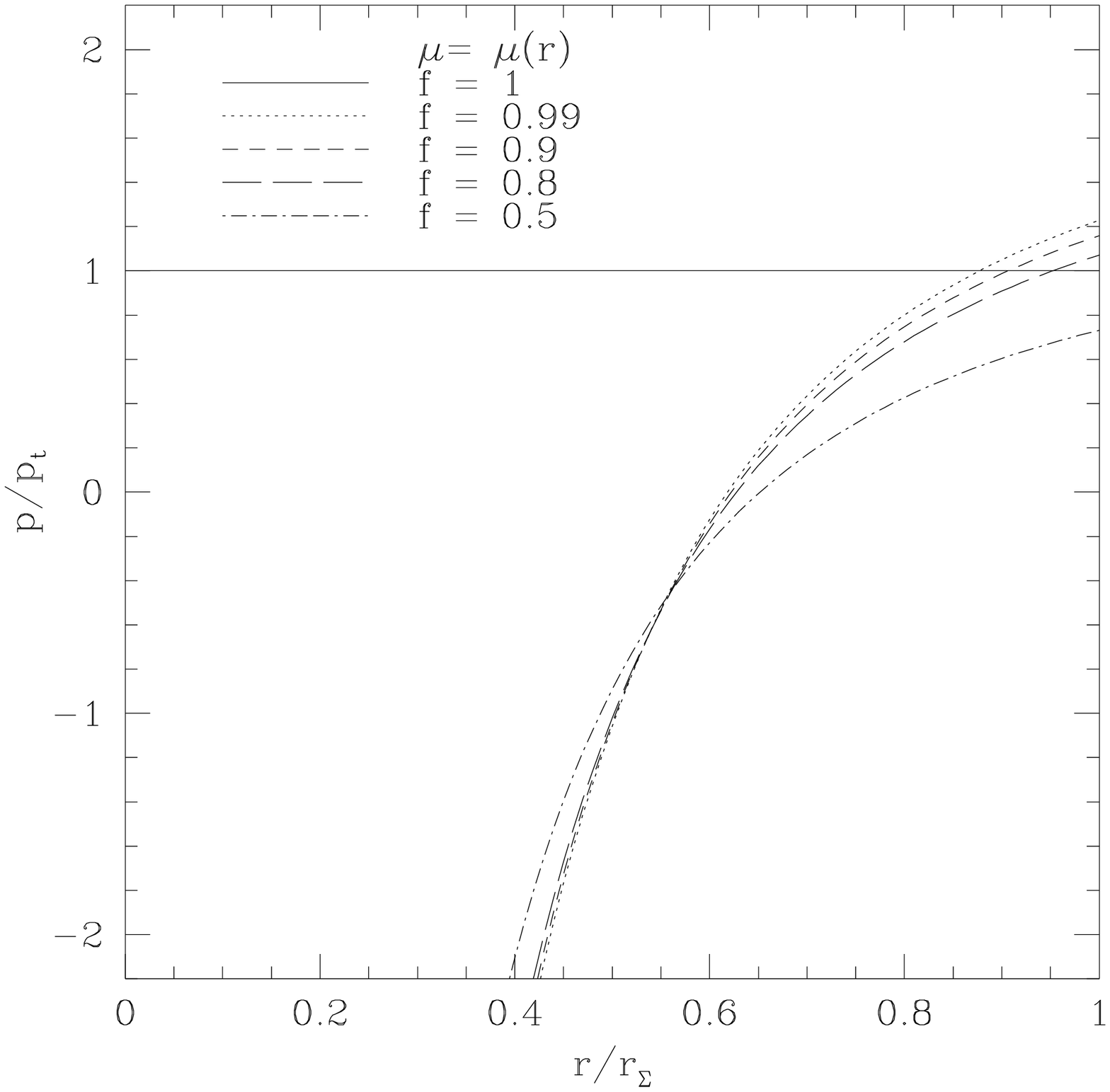,width=3.3truein,height=3.0truein}\hskip
.25in \psfig{figure=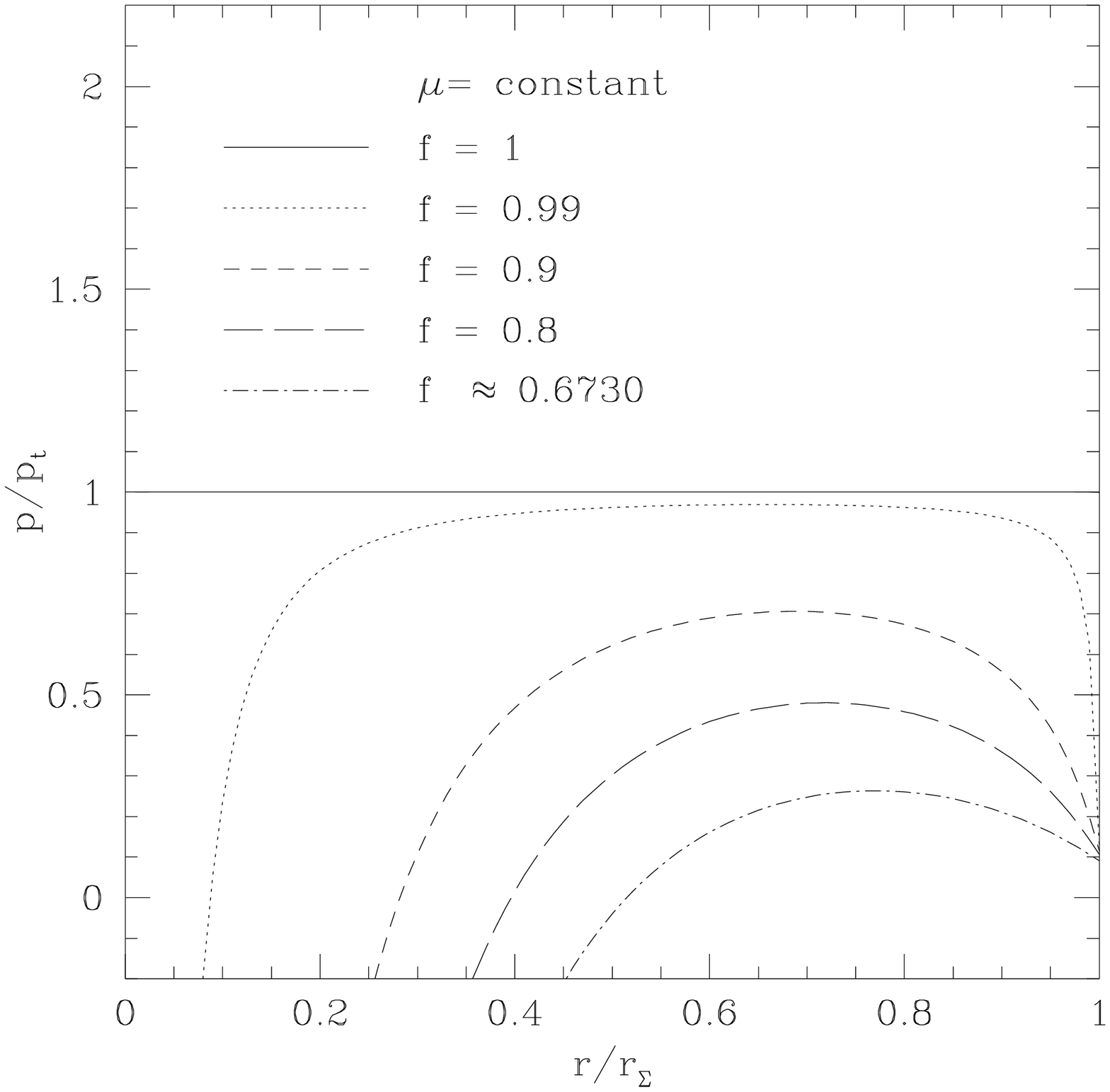,width=3.3truein,height=3.0truein}
\hskip .5in} \caption{The ratio of the radial and tangential pressure profiles 
for the model with initial homogeneous and
initial inhomogeneous energy density.  The radii $r$ and $r_{\Sigma}$ are in units of 
seconds and the radial and tangential pressures are in units of sec$^{-2}$.}
\label{prpt}
\end{figure}

\begin{figure}
\vspace{.2in}
\centerline{\psfig{figure=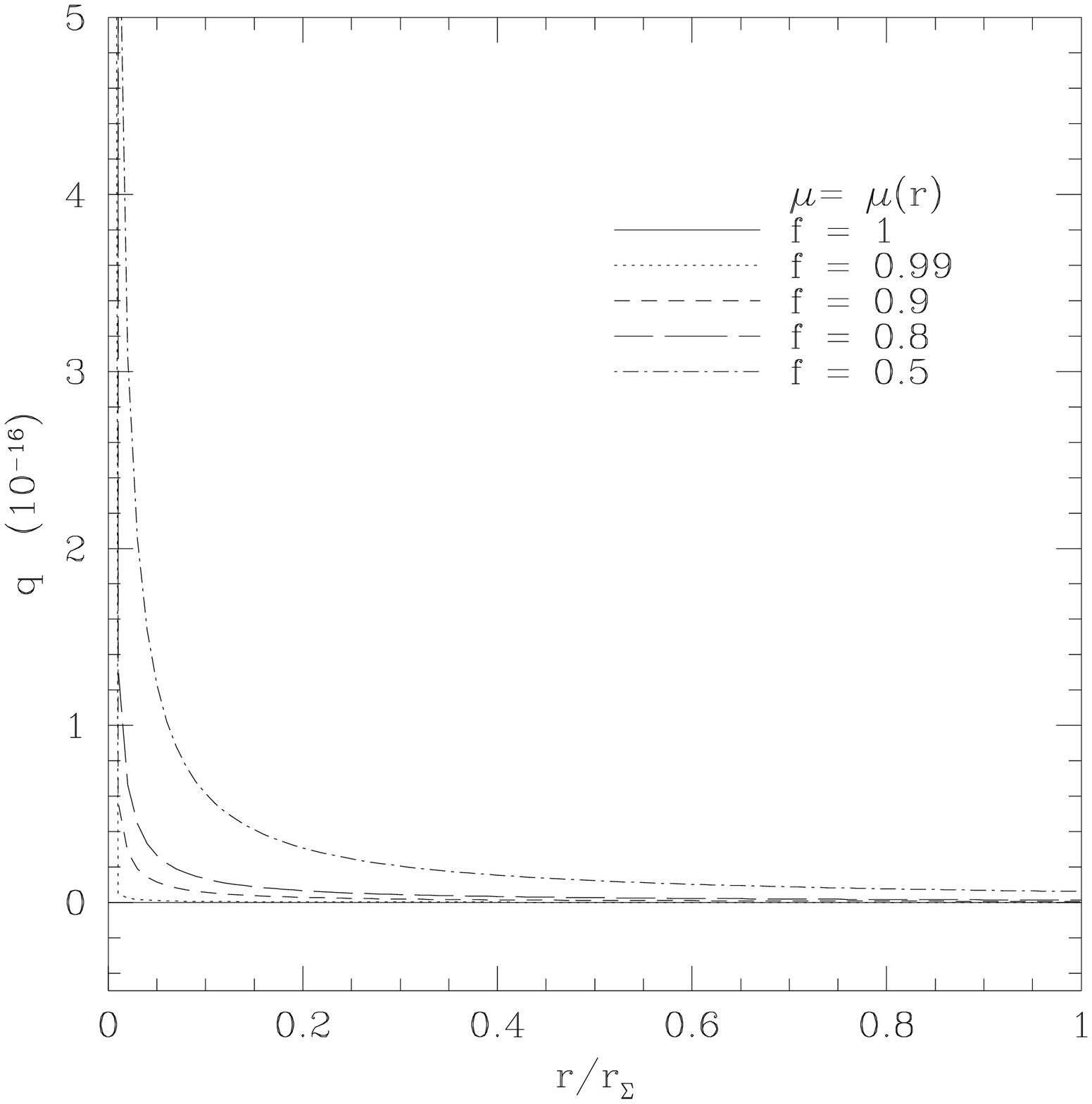,width=3.3truein,height=3.0truein}\hskip
.25in \psfig{figure=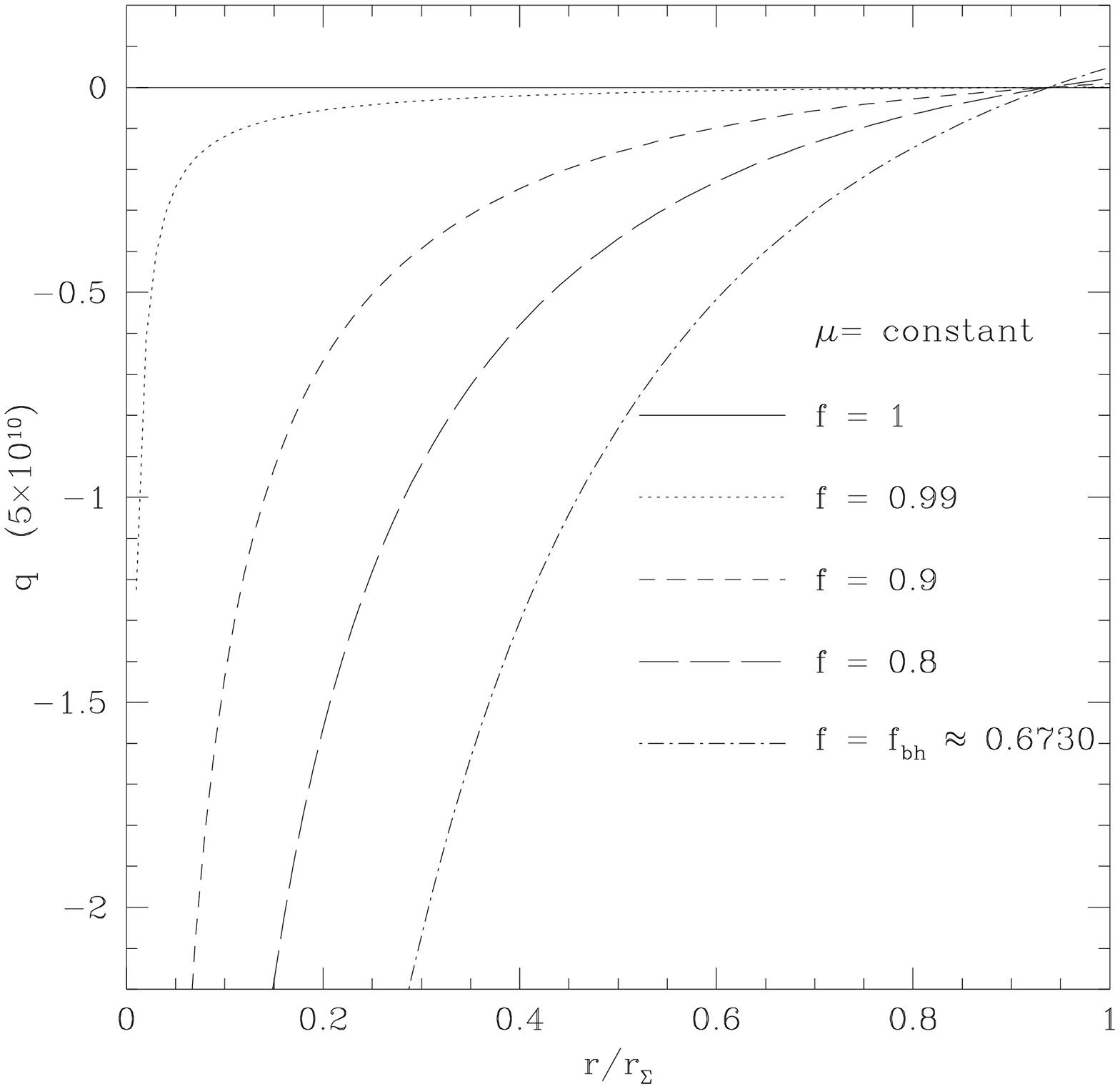,width=3.3truein,height=3.0truein}
\hskip .5in} \caption{Heat flux scalar profiles for the model with initial homogeneous and
initial inhomogeneous energy density.  The radii $r$ and $r_{\Sigma}$ are in units of 
seconds and the heat flux scalar is in units of sec$^{-2}$.}
\label{qq}
\end{figure}

It is shown in figures \ref{mu} and \ref{qq} the radial profiles of the energy density
and the heat flux.

In figures \ref{pr} and \ref{pt} we plot the radial profiles of the radial and 
tangential pressures.  

In figure \ref{prpt} is shown the radial profiles of the radial and tangential
pressure ratio.
In this figure we can see that the star is isotropic
at the beginning of the collapse ($f = 1$) but becoming more and more
anisotropic at later times.

\section{Energy Conditions for an Initial Inhomogeneous Energy Density Fluid}

All known forms of matter obey the weak, dominant and strong energy conditions.
For this reason a star model based on some fluid which violates these 
conditions cannot be seriously considered as physically relevant.

Thus, in order to find the energy conditions, we have 
followed the same procedure used in Kolassis, Santos \& Tsoubelis (1988) and
have generalized the energy conditions for an anisotropic fluid. See more
details in \cite{Chan00}\cite{Chan01}.
These conditions are fulfilled if the following inequalities are satisfied:

\begin{equation}
(i)~~~~~~ |\mu+p|-2|\bar q| \ge 0,
\label{eq:gcr1}
\end{equation}

\begin{equation}
(ii)~~~~~ \mu-p+2p_t+\Delta \ge 0,
\label{eq:gcr3}
\end{equation}

\noindent and besides,
\bigskip

\noindent a) for the weak energy conditions

\begin{equation}
(iii)~~~~ \mu-p+\Delta \ge 0,
\label{eq:wcr1}
\end{equation}

\noindent b) for the dominant energy conditions

\begin{equation}
(iv)~ \mu-p \ge 0,
\label{eq:dcr1}
\end{equation}

\begin{equation}
(v)~~~~~~ \mu-p-2p_t+\Delta \ge 0,
\label{eq:dcr2}
\end{equation}

\noindent c) for the strong energy conditions

\begin{equation}
(vi)~~~~~ 2p_t+\Delta \ge 0,
\label{eq:scr1}
\end{equation}
where $\Delta=\sqrt{(\mu+p)^2-4\bar q^2}$.

\begin{figure}
\vspace{.2in}
\centerline{\psfig{figure=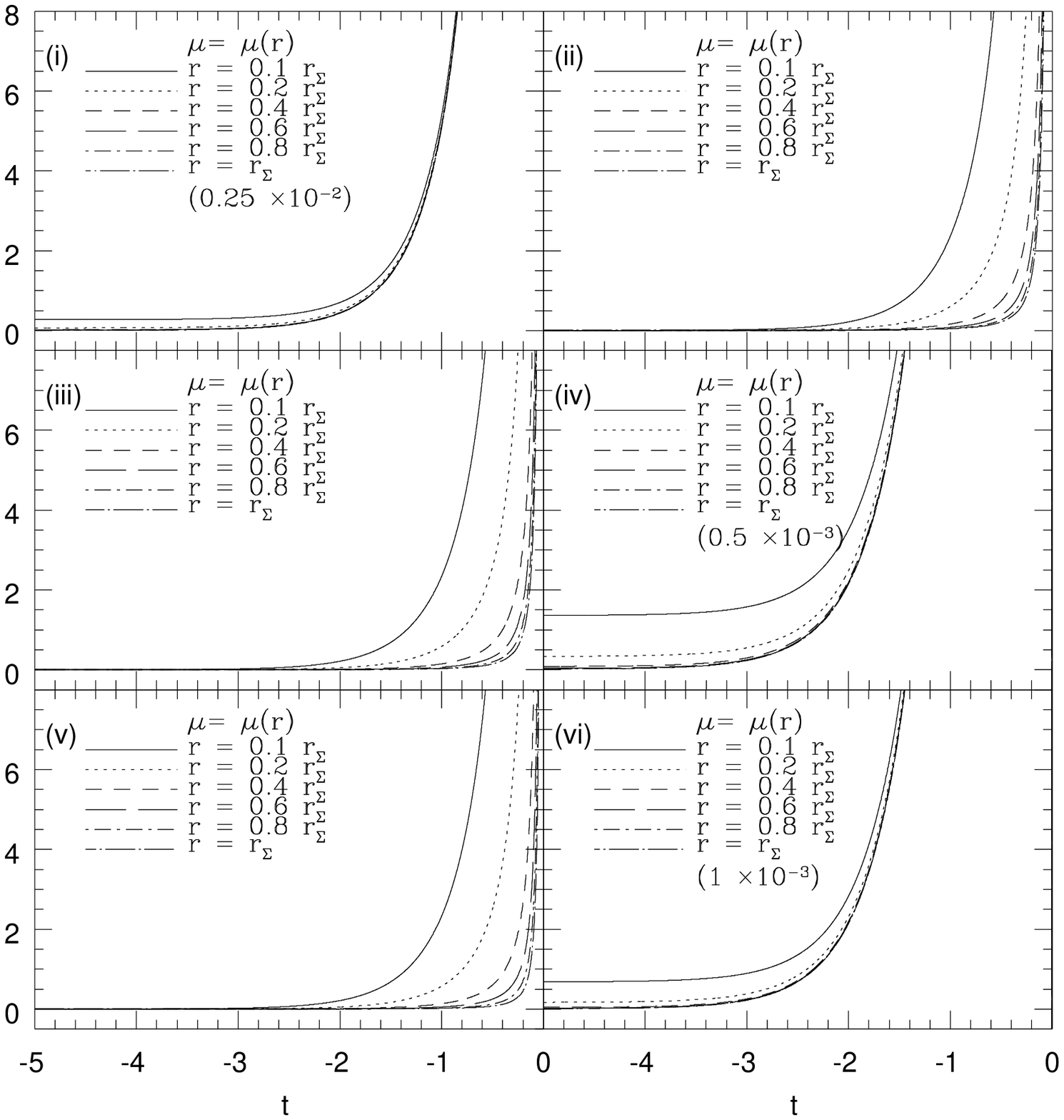,width=3.3truein,height=3.0truein}\hskip
.25in \psfig{figure=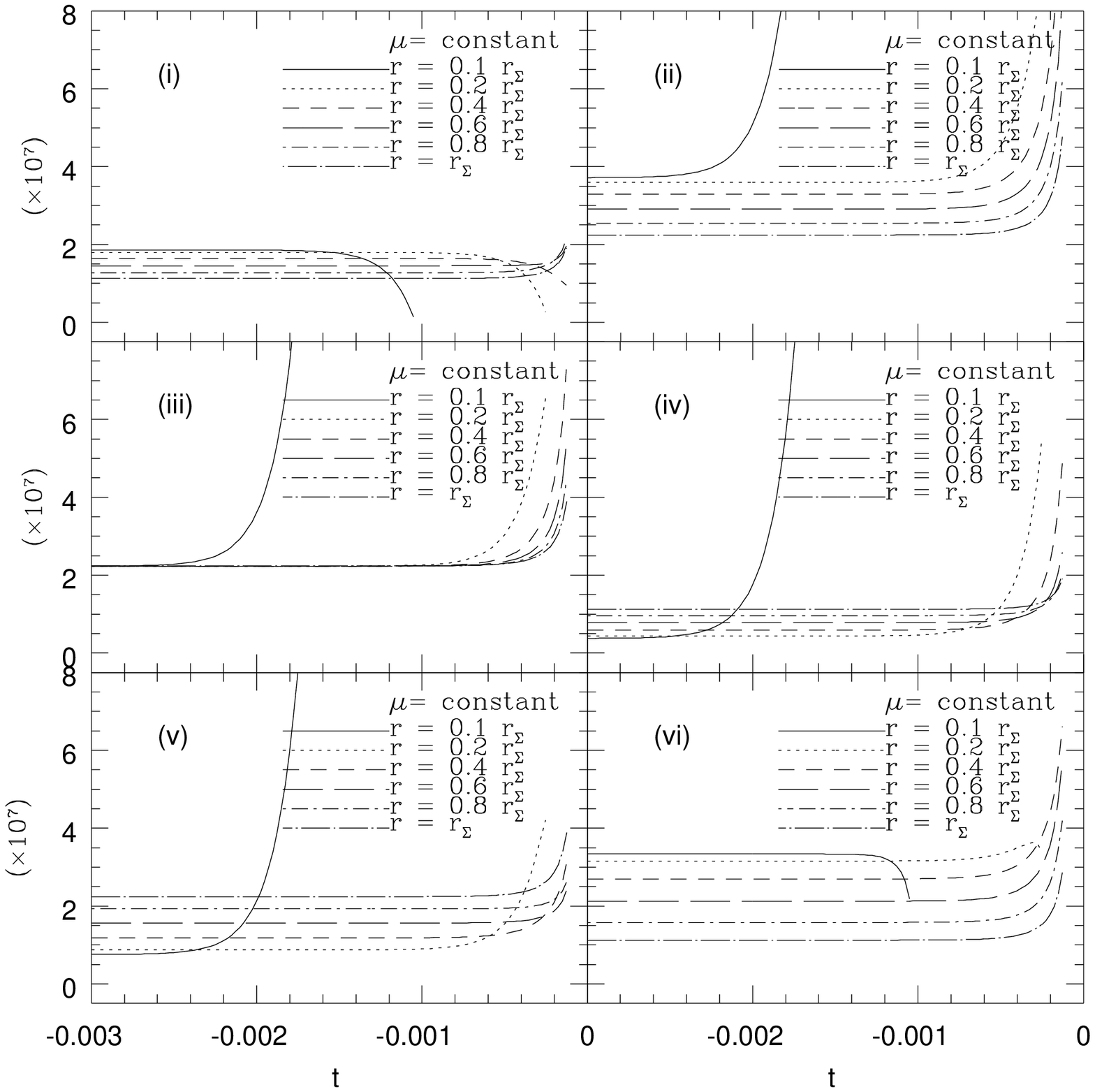,width=3.3truein,height=3.0truein}
\hskip .5in} \caption{The energy conditions (\ref{eq:gcr1})-(\ref{eq:wcr1}), for
the model with initial inhomogeneous and homogeneous energy density.
The time is in units of seconds and all the
others quantities are in units of sec$^{-2}$.}
\label{i-vi}
\end{figure}

From figure \ref{i-vi}(i)-(vi) ($\mu=\mu(r)$) we can note that the 
conditions (\ref{eq:gcr1})-(\ref{eq:scr1}) are always satisfied for all radii and times.
However, in contrast, from the figure \ref{i-vi}(i)-(vi) ($\mu=constant$) we can conclude that 
the conditions $(i)$ and $(vi)$ are not satisfied during all
the collapse and for any radius.  These inequalities are not satisfied for the 
innermost radii ($r \le 0.2r_{\Sigma}$) and for the latest stages of the 
collapse.

\section{Physical Results}

As in \cite{Chan97}\cite{Chan98a}\cite{Chan00}\cite{Chan01}, 
we have calculated several physical quantities, 
as the total energy entrapped 
inside the $\Sigma$ surface, the total luminosity perceived by an observer 
at rest at infinity and the effective adiabatic index, and we have compared
them to the respective initial homogeneous ones.

\begin{figure}
\vspace{.2in}
\centerline{\psfig{figure=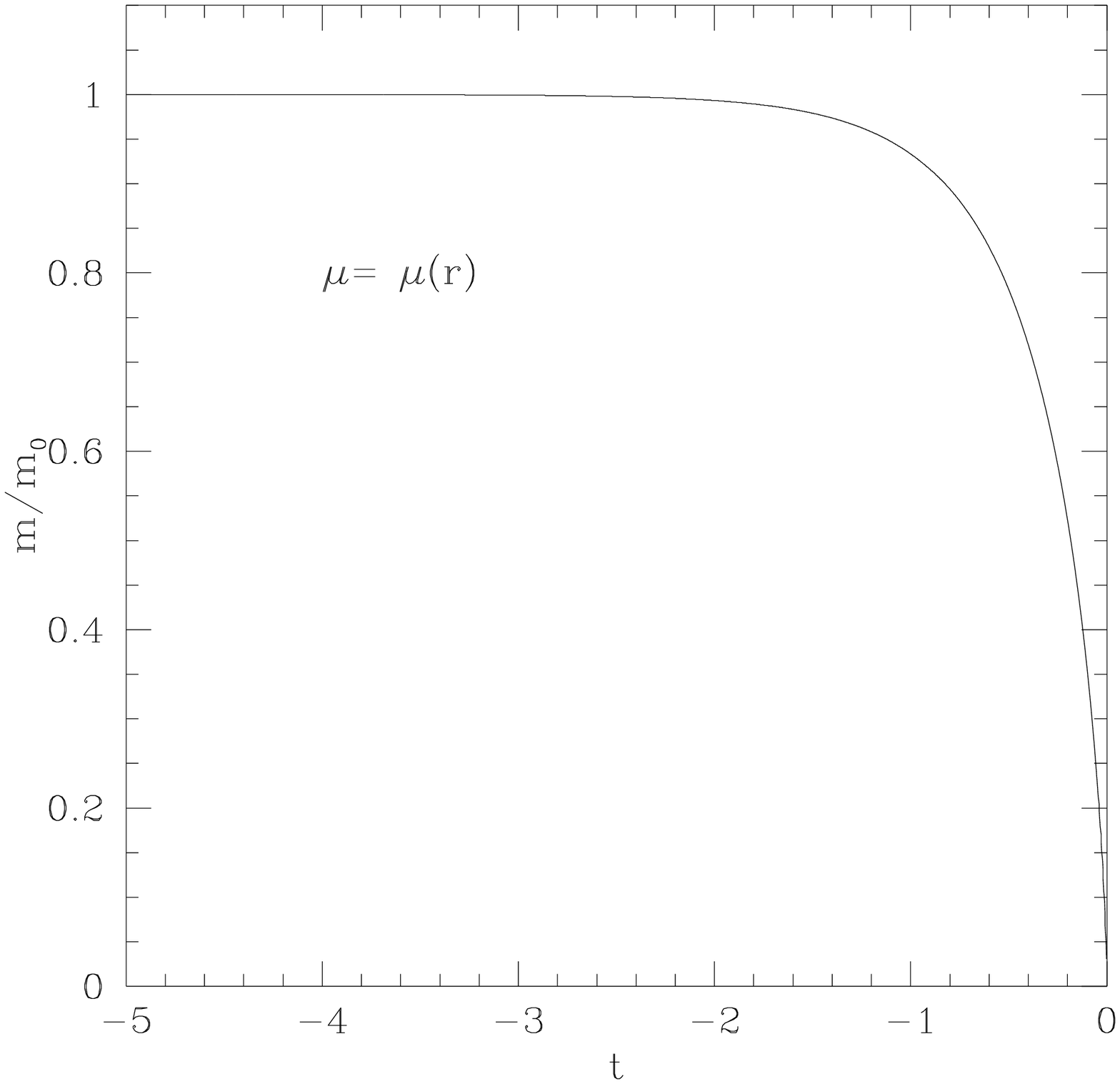,width=3.3truein,height=3.0truein}\hskip
.25in \psfig{figure=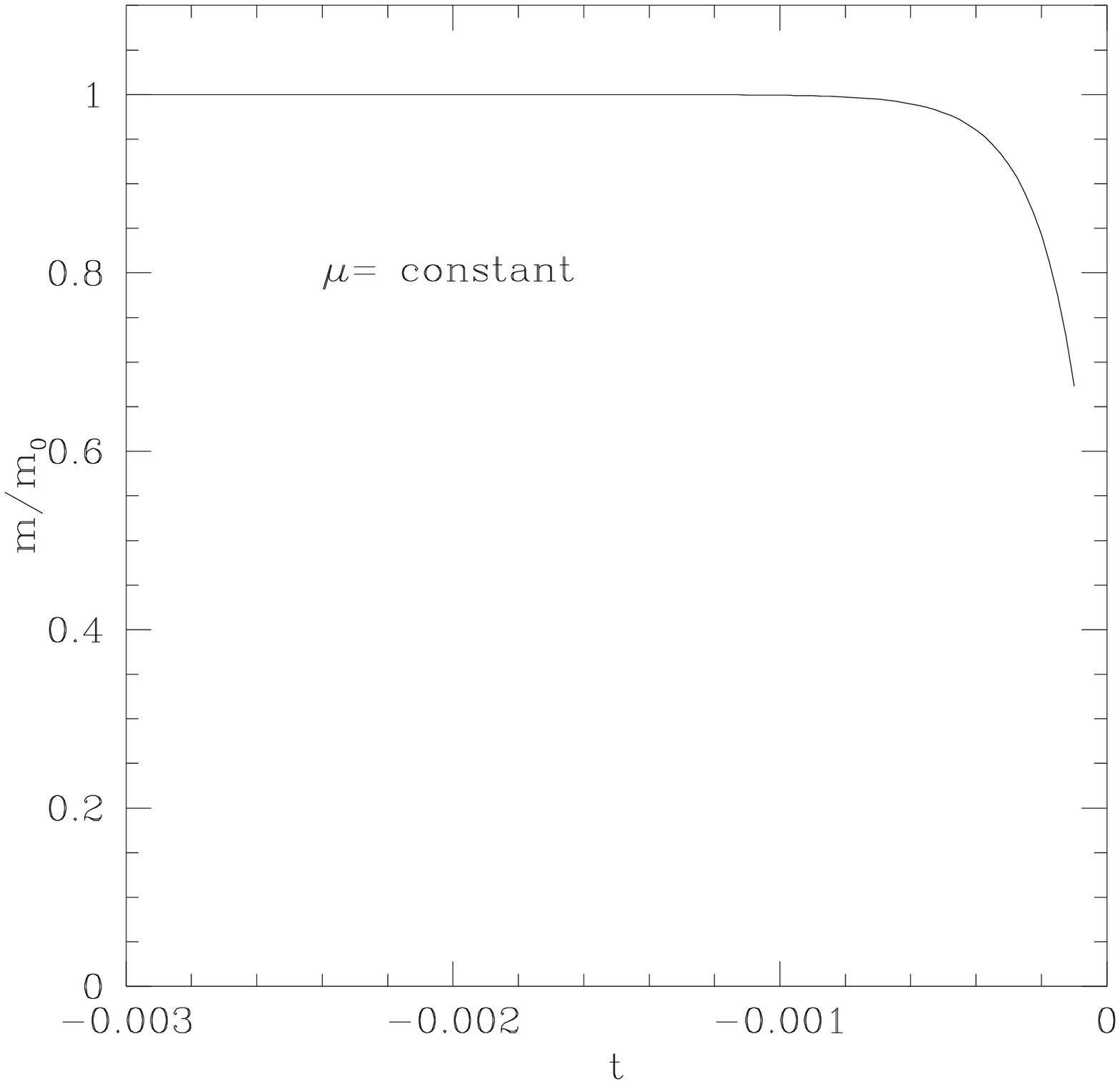,width=3.3truein,height=3.0truein}
\hskip .5in} \caption{Time behavior of the total energy entrapped inside
the surface $\Sigma$ for the models with initial inhomogeneous and homogeneous energy density.
The time, $m$ and $m_0$ are in units of seconds.}
\label{mass}
\end{figure}

From equation (\ref{eq:ms}) we can write using (\ref{eq:art0})-(\ref{eq:crt0})
that

\begin{equation}
m=\left[ {{r^3B^3_0} \over {2A^2_0}}f{\dot f}^2 + {rB_0 \over 2}
f(1 - f^2) + m_0f^3 \right]_{\Sigma},
\label{eq:mf}
\end{equation}
where

\begin{equation}
m_0=-\left[ {r^2B'_0} + {r^3B'^2_0 \over {2B_0}} \right]_{\Sigma}.
\label{eq:m0}
\end{equation}
We can observe from figure \ref{mass} that the mass inside $\Sigma$ is
different for both models, initial inhomogeneous and homogeneous energy density ones.  
The initial inhomogeneous star model radiates all its mass during the evolution, in
contrast to the initial homogeneous star model where the star radiates about 33\% of
its total mass.

\begin{figure}
\vspace{.2in}
\centerline{\psfig{figure=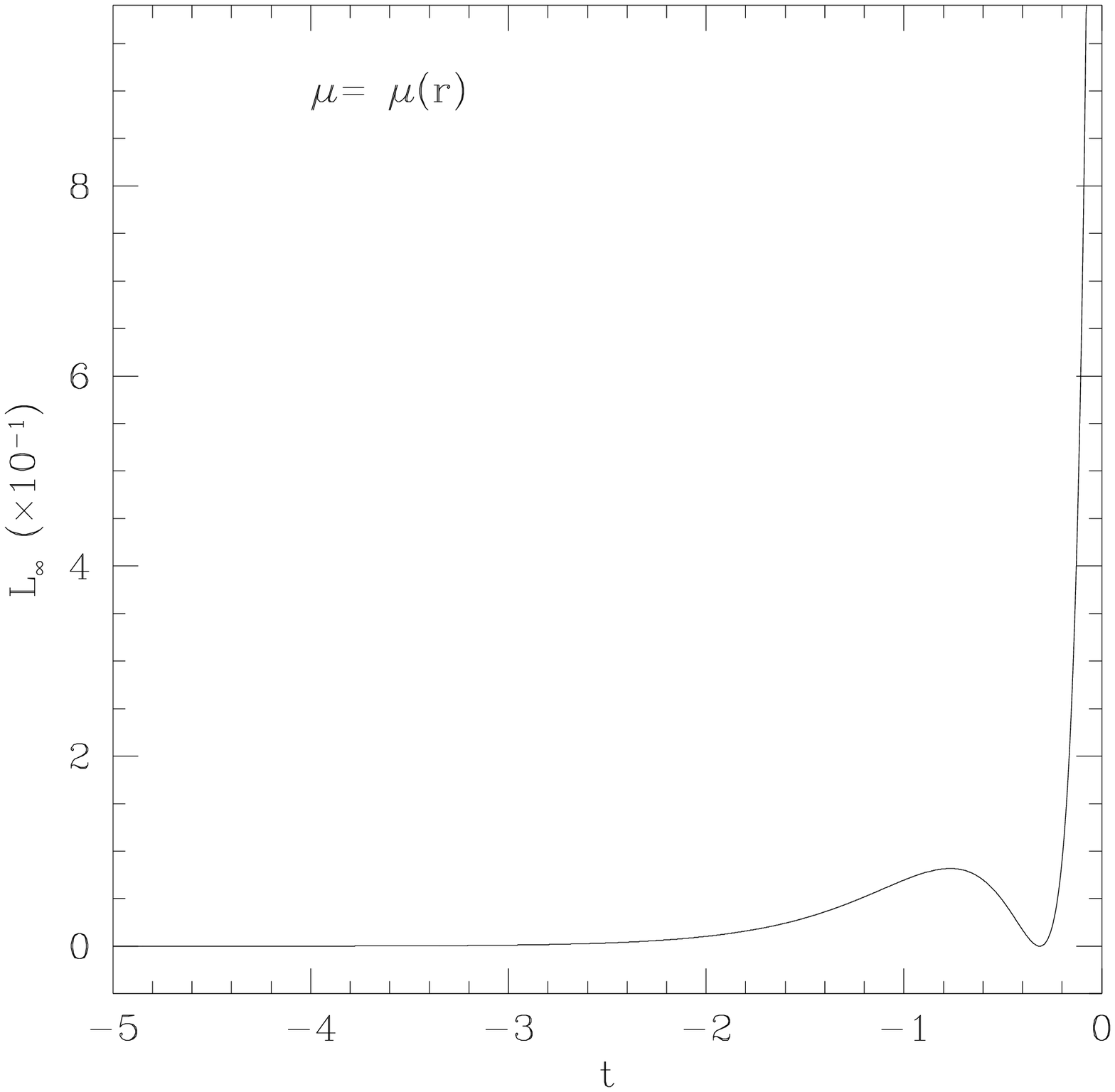,width=3.3truein,height=3.0truein}\hskip
.25in \psfig{figure=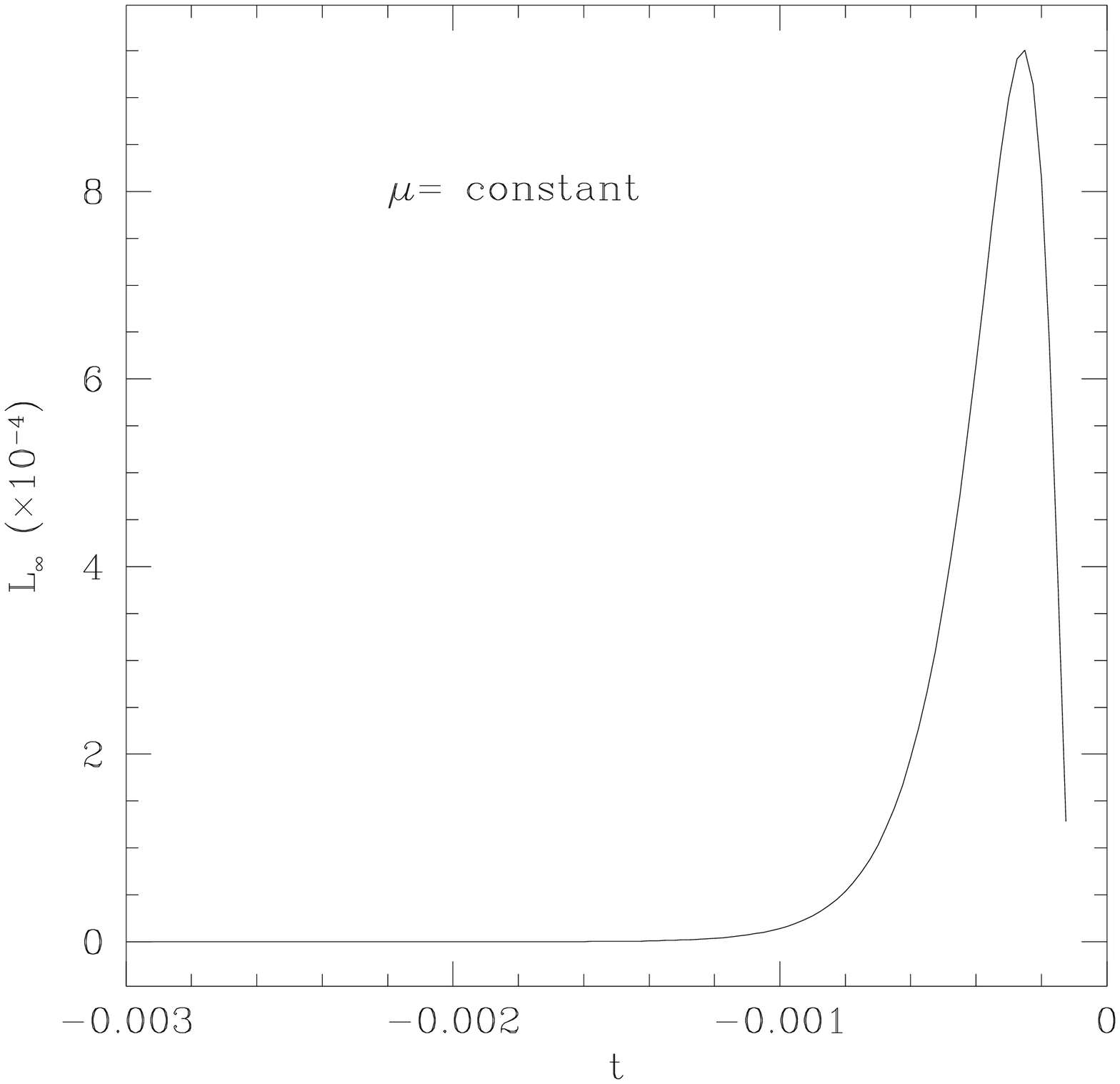,width=3.3truein,height=3.0truein}
\hskip .5in} \caption{Time behavior of the luminosity perceived by an
observer at rest at infinity for the models with initial inhomogeneous and homogeneous energy density.
The time is in units of second and the luminosity is dimensionless.}
\label{lumin}
\end{figure}

Using the equations (\ref{eq:lsp}) and (\ref{eq:art0})-(\ref{eq:crt0}) we can
write the luminosity of the star as

\begin{eqnarray}
L_{\infty}&=&{\kappa \over 2}\left\{ p r^2B^2_0f^2 \times \right. \nonumber \\ 
& & \left. \times \left[ \left( r {B'_0 \over B_0} +1 \right)f + 
\left( rB_0 \over A_0 \right) {\dot f} \right]^2 \right\}_{\Sigma}.
\label{eq:lf}
\end{eqnarray}

The effective adiabatic index can be calculated using the equations
(\ref{eq:mu})-(\ref{eq:p}), (\ref{eq:ft}) and 
(\ref{eq:a0})-(\ref{eq:p0a}). Thus, we can write that 

\begin{eqnarray}
\Gamma_{\rm eff}=
\left[ {\partial (\ln p)} \over {\partial (\ln \mu)} \right]_{r=const} 
= \left( {\dot p} \over p \right) \left( \mu \over {\dot \mu} \right) & = &\nonumber \\
\frac{aB_0^2f \left[ -a f \dot f  - b(1-f^2) -3 \dot f^2 \right] - 4(b B_0^2 - A_0^2/r^2) \dot f}
{f^2 A_0^2 B_0^2 \kappa p_0 + a B_0^2 f \dot f + (B_0^2 b - A_0^2/r^2)(1- f^2)} \times \nonumber \\
\frac{f^2 A_0^2 B_0^2 \kappa \mu_0 + B_0^2 \dot f^2 + A_0^2(1-f^2)/r^2}
{2B_0^2 \dot f \left[ -a f \dot f - b(1-f^2) - 3 \dot f^2 \right] -4 \dot f/A_0^2}.
\label{eq:gamaf}
\end{eqnarray}

\begin{figure}
\vspace{.2in}
\centerline{\psfig{figure=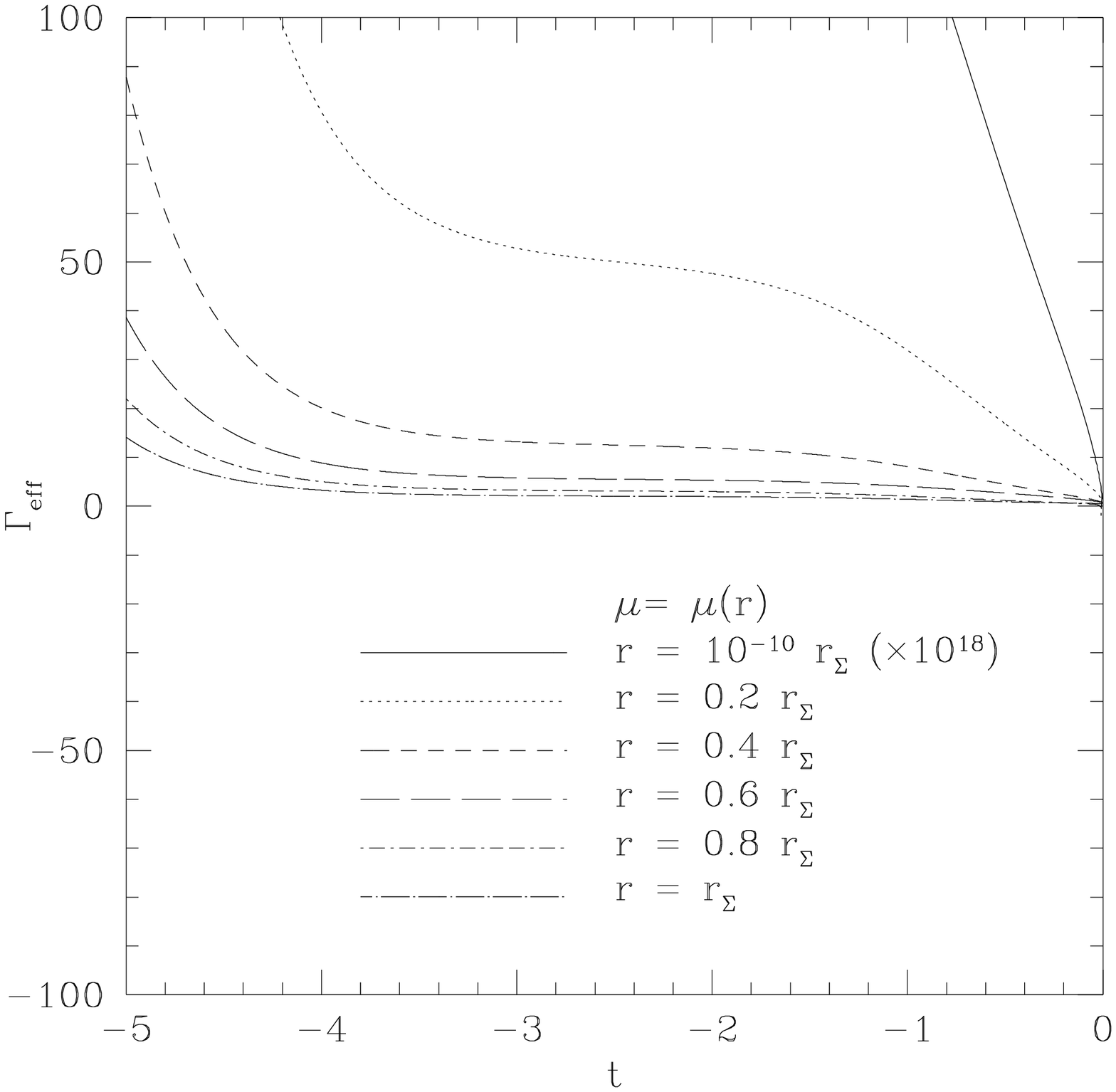,width=3.3truein,height=3.0truein}\hskip
.25in \psfig{figure=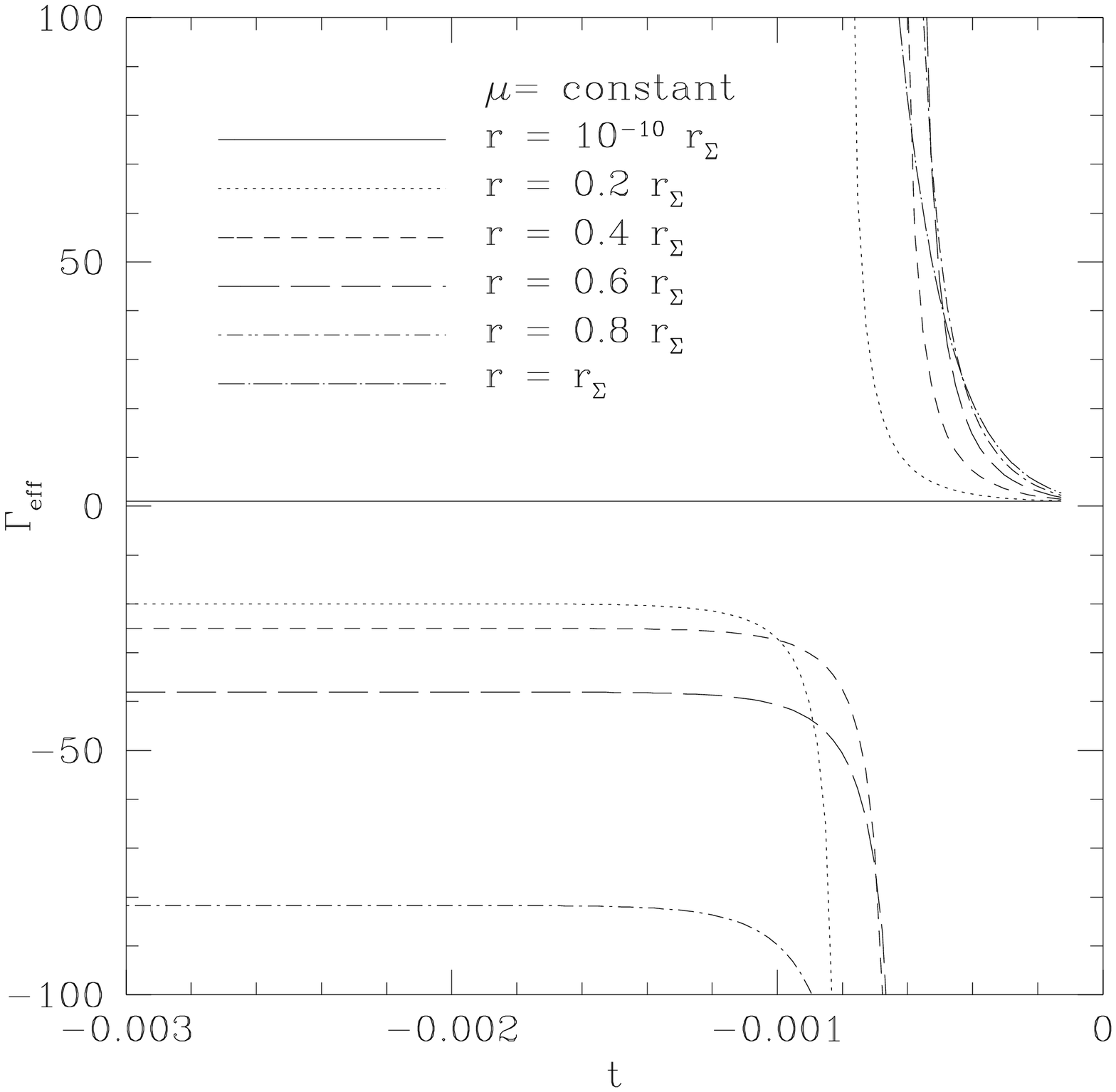,width=3.3truein,height=3.0truein}
\hskip .5in} \caption{Time behavior of the effective adiabatic index
for the models with initial inhomogeneous and homogeneous energy density.
The time is in units of second and $\Gamma_{\rm eff}$ is dimensionless.}
\label{gama}
\end{figure}

Comparing the figures for $\Gamma_{\rm eff}$ ($\mu = \mu(r)$ and $\mu = constant$) 
we notice that the time  
evolution of the effective adiabatic indices are very different graphically.
We can note in figure \ref{gama} ($\mu = constant$) that shortly before the 
peak of luminosity (see figure \ref{lumin}) there is a large discontinuity
in $\Gamma_{\rm eff}$, in contrast to the initial inhomogeneous star model where there
is no discontinuity in $\Gamma_{\rm eff}$.

\section{Conclusions}

A new model is proposed to a collapsing star consisting of an
initial inhomogeneous energy density fluid, radial heat flow and outgoing
radiation.  In previous papers \cite{Chan97}-\cite{Chan01} one of us has
introduced a collapse model of an initial homogeneous energy density.
In this work we have generalized
this previous model by introducing an initial inhomogeneous energy
density and we have compared it to the initial homogeneous one.

The behavior of the energy density, pressure, mass, luminosity and the
effective adiabatic index was analyzed. We have also compared to the case
of a collapsing fluid with initial homogeneous energy density of another previous model \cite{Chan97}\cite{Chan98a},
for a star with 6 $M_{\odot}$.

As we have shown the black hole is never formed
because the apparent horizon formation condition is never satisfied.
This could be interpreted as the formation of a naked singularity, as
Joshi, Dadhich and Maartens \cite{Joshi02} have suggested.  However
this is not the case because the star radiates all its mass before it
reaches the singularity at $r=0$ and $t=0$.  Not even a marginally naked
singularity is formed by the same reason, since in this case the apparent
horizon should coincide with the singularity at $r=0$ and $t=0$.

The pressure has negative values although physically this
could be considered unreasonable.  However, due to the heat flow (the term
$\Delta$) the energy conditions are always satisfied.

The pressure of the star, at the beginning of the collapse, is isotropic but
due to the presence of the shear the pressure becomes more and more
anisotropic.

The star radiates all its mass during the collapse and this explains
why the apparent horizon never forms.
In contrast of the result of this work, the
former model radiates about 33\% of the total mass of the star, before the
formation of the black hole.

An observer at infinity will see a radial point source radiating
exponentially until reaches the time of maximum luminosity and
suddenly the star turns off because there is no more mass in order to be
radiated.
In contrast of the former model with initial homogeneous energy density \cite{Chan97}-\cite{Chan01}
where the luminosity also increases exponentially, reaching a maximum and after
it decreases until the formation of the black hole.

The effective adiabatic index, in the initial inhomogeneous energy density model,
is always positive without any discontinuity in contrast of the former model
where there is a discontinuity around the time of maximum luminosity.
In the case of initial homogeneous energy density,
the effective adiabatic index has a very unusual behavior because we have
a non-adiabatic regime in the fluid due to the heat flow.  The index becomes
negative since the hydrodynamic pressure and the energy density may become negative.
Besides, in this case, neither the energy density is the measure of the
total energy density of a given piece of matter nor the hydrodynamic
pressure the only opposing contraction \cite{Barreto92}.

\bigskip
\bigskip

\noindent{\bf ACKNOWLEDGMENTS}

The author (RC) acknowledges the financial
support from FAPERJ (no. E-26/171.754/2000, E-26/171.533/2002 and
E-26/170.951/2006) and from
Conselho Nacional de Desenvolvimento Cient\'{\i}fico e Tecnol\'ogico - CNPq -
Brazil.
The author (GP) also acknowledges the financial support from CAPES.

\end{document}